\documentclass[english,12pt,aps,prd,a4paper,preprintnumbers, floatfix,nofootinbib,showpacs,superscriptaddress, notitlepage]{revtex4-1} 
 \pdfoutput=1
\usepackage[usenames,dvipsnames]{color}  
\usepackage{graphicx}
\usepackage{setspace}
\usepackage{braket}
\usepackage{caption}
\usepackage{subcaption}
\usepackage{amsmath}
\usepackage{amssymb}
\usepackage[colorlinks=true,citecolor=darkred,urlcolor=darkred, pdfborder={0 0 0}]{hyperref}
\usepackage[normalem]{ulem}
\usepackage{xcolor}
\usepackage{graphicx}
\makeatletter
\def\p@subsection{}
\makeatother
\usepackage{xcolor}
\usepackage{float}

\definecolor{darkred}{rgb}{0.6,0,0}

\definecolor{linkcolor}{rgb}{0,0,0.5}

\def\gsim{\raise0.3ex\hbox{$\;>$\kern-0.75em\raise-1.1ex\hbox{$\sim\;$}}}
\def\lsim{\raise0.3ex\hbox{$\;<$\kern-0.75em\raise-1.1ex\hbox{$\sim\;$}}}

\def\beqn#1{\begin{equation}\label{#1}}
\def\eeqn{\end{equation}}

\def\beqa#1{\begin{eqnarray}\label{#1}}
\def\eeqa{\end{eqnarray}}

\usepackage{caption}
\usepackage{subcaption}
\usepackage{adjustbox}

\def\Z2{$\mathcal{Z_2}$}

\usepackage{bbold}
\usepackage{ragged2e}

\newcommand {\ignore}[1]{}

\def\cevns{CE$\nu$NS }
\def\eves{E$\nu$ES }
\usepackage{mathrsfs}

\def\321{$\mathrm{SU(3) \otimes SU(2) \otimes U(1)}$ }

\newcommand{\AddrIISERB}{Department of Physics, Indian Institute of Science Education and Research - Bhopal, \\ 
Bhopal Bypass Road, Bhauri, Bhopal 462066, India}

\newcommand{\AddrIoannina}{%
Division of Theoretical Physics, University of  Ioannina, GR 45110 Ioannina, Greece}

\begin{document}

\title{\textcolor{BrickRed}{Dark matter detectors as a novel probe for light new physics}}

\author{Anirban Majumdar}\email{anirban19@iiserb.ac.in}
\affiliation{\AddrIISERB}
\author{D. K. Papoulias}\email{d.papoulias@uoi.gr}
\affiliation{\AddrIoannina}
\author{Rahul Srivastava}\email{rahul@iiserb.ac.in}
\affiliation{\AddrIISERB}

\begin{abstract}

We explore the prospect of constraining light mediators at the next generation direct detection dark matter detectors through coherent elastic neutrino-nucleus scattering (CE$\nu$NS) and elastic neutrino-electron scattering (E$\nu$ES) measurements. Taking into account various details like the quenching factor corrections, atomic binding effects, realistic backgrounds, detection efficiency, energy resolution etc., we consider two representative scenarios regarding detector specifications. For both scenarios, we obtain the  model-independent projected sensitivities for all possible  {Lorentz-invariant} interactions, namely scalar ($S$), pseudoscalar ($P$), vector ($V$), axial vector ($A$) and tensor ($T$). For the case of vector interactions, we also  focus on two concrete examples: the well-known $U(1)_{B-L}$ and $U(1)_{L_\mu - L_\tau}$ gauge symmetries. For all interaction channels $X=\{S,P,V,A,T\}$, our results imply that the upcoming dark matter detectors have the potential to place competitive constraints, improved by about  {1} order of magnitude compared to existing ones from dedicated CE$\nu$NS experiments, XENON1T, beam dump experiments and collider probes.

\end{abstract}
\maketitle

\section{Introduction}

Although the Standard Model (SM) provides a rather successful description of electroweak and strong interactions in nature, there is a list of shortcomings that points to the need of new physics  {beyond} the Standard Model (BSM).
Usually it is assumed that the scale of such new physics is larger than the electroweak symmetry breaking scale and the new BSM particles are heavy. For example, the existence of new neutral $Z^\prime$ gauge bosons under an extra gauge symmetry is a common feature of many theories beyond the SM~\cite{Langacker:2008yv}. Extensive phenomenological studies at colliders~\cite{Dittmar:2003ir} have been performed, taking the  $Z'$ bosons to be heavier than electroweak scale.

However, there exist various motivated BSM extensions where the new physics is desirable to be at the low scale. Light new physics, for example, can account for the explanation of existing anomalies such as the longstanding anomalous magnetic moment of  {muons}~\cite{Muong-2:2021ojo,Aoyama:2020ynm,Jegerlehner:2009ry}. Moreover, concerning the dark sector, several  dark matter models require the presence of light mediators to account for dark matter  {self-interaction}~\cite{Agrawal:2021dbo,Fabbrichesi:2020wbt} and/or the recent XENON1T anomaly~\cite{Boehm:2020ltd,AristizabalSierra:2020edu, Khan:2020vaf}. Finally, the  Peccei-Quinn solutions of  {the strong $CP$ problem also imply} the presence of a light Nambu-Goldstone boson called axion~\cite{Peccei:1977hh,Weinberg:1977ma,Wilczek:1977pj},  while neutrino mass models involving dynamical lepton number breaking often lead to a light pseudoscalar boson called Majoron (Majorana  {neutrinos}) or Diracon (Dirac neutrinos)~\cite{Schechter:1981cv,Bonilla:2018ynb}.

While the implications of light vector and scalar mediators to the expected signal rates had already been explored at direct detection dark matter experiments~\cite{Cerdeno:2016sfi,Bertuzzo:2017tuf}, after the recent observation of coherent elastic neutrino-nucleus scattering (CE$\nu$NS) by the COHERENT experiment~\cite{COHERENT:2017ipa,COHERENT:2020iec,Akimov:2021dab} there has been an intense interest for novel mediator investigations~\cite{Dent:2016wcr,Liao:2017uzy, Farzan:2018gtr, Papoulias:2019txv, Dutta:2019eml, AristizabalSierra:2019ykk, Miranda:2020tif, Cadeddu:2020nbr}. In addition to COHERENT, data-driven  constraints also exist from upper limits  on CE$\nu$NS, placed by the recent CONNIE~\cite{CONNIE:2019xid} and CONUS~\cite{CONUS:2021dwh}  measurements. Phenomenological studies focusing on various $U^\prime(1)$  {realizations} such as the $L_\mu - L_\tau$ and $B-L$  {gauge} symmetries have been explored using solar neutrinos~\cite{Sadhukhan:2020etu,Amaral:2020tga,Amaral:2021rzw}, as well as using supernova neutrinos taking also into account corrections from medium effects~\cite{Cerdeno:2021cdz}. It is interesting to note that the latter can induce a reduction of the cross section due to Pauli blocking and therefore lead to less severe constraints. Very recently, the expected modification to the  {neutrino floor} has been illustrated in the presence of vector and scalar mediators~\cite{AristizabalSierra:2021kht}, while the complementarity of \cevns and direct detection dark matter experiments was  {emphasized}  through a data-driven analysis of the neutrino floor. 

In addition to scalar and vector mediator scenarios, it is possible to explore further  {Lorentz-invariant} structures, such as 
 {scalar} ($S$),  {pseudoscalar} ($P$),  {vector} ($V$),  {axial vector} ($A$), or tensor ($T$) interactions in a  {model-independent way through neutrino  generalized interactions} (NGIs). Assuming heavy mediators,  NGIs have been explored in Ref.~\cite{AristizabalSierra:2018eqm} using \cevns data from COHERENT and in Ref.~\cite{Khan:2019jvr} through elastic neutrino-electron scattering (E$\nu$ES) by  {analyzing} the Borexino data.  Reference~\cite{Escrihuela:2021mud} performed a global NGI analysis in the light data from E$\nu$ES experiments, neutrino  {deep inelastic scattering}, and single-photon detection from electron-positron collisions, while Ref.~\cite{Rodejohann:2017vup} pointed out the possibility of probing the Dirac or Majorana neutrino nature through NGIs. Tensorial exotic interactions have been explored in Ref.~\cite{Barranco:2011wx} using the TEXONO data as well as in Ref.~\cite{Papoulias:2015iga} where their connection to neutrino transition magnetic moments was  {emphasized}. Finally, constraints in the tensor parameter space were extracted from the analysis of COHERENT data~\cite{Papoulias:2017qdn} and more recently by the CONUS Collaboration~\cite{CONUS:2021dwh}.

The next generation direct detection dark matter experiments with  {multiton} mass scale and sub-keV capabilities are expected to become sensitive to astrophysical neutrinos. Indeed, the XENON1T Collaboration has already reported first results in its effort to identify a potential \cevns population--induced by $^{8}$B neutrinos of the solar flux--in  the background data~\cite{XENON:2020gfr}. 
Motivated by the upcoming large scale next generation direct detection dark matter experiments such as XENONnT~\cite{XENON:2020kmp}, in this work we consider the possibility of exploring  neutrino backgrounds using nuclear and electron recoils. 
In particular, we are interested to explore the prospects of constraining  general neutrino interactions induced by novel mediators via \cevns and \eves measurements at a future direct detection dark matter experiment. Unlike previous studies--especially for the cases of pseudoscalar, axial vector and tensor interaction channels--we allow the exchanged mediator to be sufficiently light. Therefore, for all possible interaction channels, the dependence on the mediator mass has been considered explicitly in the cross sections. Our present results indicate that future dark matter detectors will offer competitive constraints in the mass-coupling parameter space, being also complementary to collider probes and DUNE.

The remainder of the paper has been  {organized} as follows. In Sec.~\ref{sec:CEvNS} we provide the necessary formalism regarding \cevns processes. We start with the discussion of  \cevns within  {the} SM and then discuss the new physics contribution to it. In Sec.~\ref{sec:EveS} we discuss the case of E$\nu$ES, again starting with the SM discussion and ending with the new physics contribution. Our main results are presented in Sec.~\ref{sec:results} where we first discuss the  model-independent constraints on various  types of new physics scenarios. We then, as examples,  take a few specific models with light mediators and further  {analyze} our constraints, comparing and contrasting them with constraints obtained from other experimental probes.  Our concluding remarks are  {summarized} in Sec.~\ref{sec:conclusions}.

\section{Coherent Elastic Neutrino-Nucleus Scattering}
\label{sec:CEvNS}

In this section, we provide the basic formalism for the description of the various \cevns interaction channels considered in the present work, within and beyond the SM. A  model-independent analysis is performed by considering all possible  Lorentz-invariant interaction channels in the  {low-energy} regime. The latter can be conveniently classified in terms of the Lorentz and  {parity} transformations of the mediator particle i.e. whether the mediator transforms as $S,P,V,A$, or $T$. One further advantage of such   {parametrization} is that interactions of any particle having a mixed transformation (e.g. SM $Z^0$ boson) can also be taken into account simply as a combination of $S,P,V,A,T$ interactions.

\subsection{CE$\nu$NS within the SM}
\label{subsec:CEvNS_SM}
Assuming SM interactions only, for low and intermediate neutrino energies $(E_\nu \ll M_{Z^0})$ \cevns is accurately  described in the context of an effective  {four-fermion} Fermi interaction Lagrangian~\cite{Barranco:2005yy}
 \begin{equation}
 \label{equn:CEvNS_SM_Lagrangian}
 \mathscr{L}_{SM}=-2\sqrt{2}G_F\sum_{\substack{f=u,d\\ \alpha=e,\mu,\tau}}g_{\alpha,\alpha}^{f,P}\left[\bar{\nu}_\alpha\gamma^\rho L \nu_\alpha\right]\left[\bar{f}\gamma_\rho P f\right]\, .
 \end{equation}
Here, $P\equiv \{L,R\}$ stand for the chiral projection operators, $f\equiv \{u,d\}$ represents the first generation quark and $g_{\alpha,\alpha}^{f,P}$ is the $P$-handed coupling of the quark $f$ to the SM $Z^0$ boson. The latter are expressed in terms of the weak mixing angle $(\sin^2\theta_W=0.2387)$, as 
\begin{equation}
\begin{aligned}
g_{\alpha,\alpha}^{u,L} & = \frac{1}{2}-\frac{2}{3}\sin^2\theta_W\, ,
\qquad ~~g_{\alpha,\alpha}^{u,R} =  -\frac{2}{3}\sin^2\theta_W \, , \\ 
 g_{\alpha,\alpha}^{d,L} & =  -\frac{1}{2}+\frac{1}{3}\sin^2\theta_W\, ,
\qquad  g_{\alpha,\alpha}^{d,R}   =  \frac{1}{3}\sin^2\theta_W \, .
\end{aligned}
\end{equation}
At tree level\footnote{Subdominant radiative corrections are discussed in Ref.~\cite{Tomalak:2020zfh}.}, the SM differential \cevns cross section with respect to the nuclear recoil energy $E_{nr}$  is given as \cite{AristizabalSierra:2020zod}
\begin{equation}
\label{equn:CEvNS_SM_xsec}
\left[\frac{d\sigma}{dE_{nr}}\right]_{SM}^{\nu N}=\frac{G_F^2 m_N}{\pi}\left({Q_V^{SM}}\right)^2\left(1-\frac{m_N E_{nr}}{2E_\nu^2}\right) \, ,
\end{equation}
where $G_F$ is the Fermi constant and $m_N$ is the nuclear mass, while the SM vector weak charge $Q_V^{SM}$ takes the form \cite{Papoulias:2018uzy}
\begin{equation}
\label{equn:CEvNS_SM_charges}
Q_V^{SM}=\left[g_p^V Z+g_n^V N\right]F_W(q^2) \, .
\end{equation}
Here, $Z$ and $N$ denote the number of protons and neutrons in the nucleus,  while the corresponding vector couplings for protons $(g_p^V)$ and neutrons $(g_n^V)$ read~\cite{Papoulias:2015vxa}
\begin{equation}
\begin{aligned}
g_{p}^V=& 2\left(g_{\alpha,\alpha}^{u,L}+g_{\alpha,\alpha}^{u,R}\right)+\left(g_{\alpha,\alpha}^{d,L}+ g_{\alpha,\alpha}^{d,R} \right)= 1/2 -2\sin^2{\theta_W}\, , \\
g_n^V=& \left(g_{\alpha,\alpha}^{u,L}+g_{\alpha,\alpha}^{u,R}\right)+2\left(g_{\alpha,\alpha}^{d,L}+ g_{\alpha,\alpha}^{d,R} \right)=- 1/2 \, .
\end{aligned}
\label{equn:CEvNS_SM_nuclear_couplings}
\end{equation}
It is noteworthy that Eq.(\ref{equn:CEvNS_SM_xsec}) is valid for sufficiently low momentum transfer in order to satisfy the coherency condition $q \leq 1/R_A$~\cite{Papoulias:2018uzy}, with $R_A$ being the nuclear radius and $q^2=2m_N E_{nr}$ denoting the magnitude of  {three-momentum} transfer. Moreover, to account for the finite nuclear spatial distribution, nuclear physics corrections are incorporated in Eq.(\ref{equn:CEvNS_SM_charges}) through the weak nuclear form factor $F_W(q^2)$. In our present study we  adopt the Helm parametrization, given as~\cite{Helm:1956zz}
\begin{equation}
\label{equn:Helm}
F_W(q^2)=\frac{3j_1(qR_0)}{q R_0} \,  e^{\left[-\frac{1}{2}(qs)^2\right]} \, ,
\end{equation}
where $j_1(x)$ is  {the first-order spherical} Bessel function, while the diffraction radius is given by $R_0^2=\frac{5}{3}R^2-5s^2$ with the nuclear radius and surface thickness taken to be $R=1.23\, A^{1/3}$~fm and $s=0.9$~fm,  respectively with $A=N+Z$ being the atomic mass number.

\vspace*{-1.46cm}

\subsection{CE$\nu$NS contribution from light novel mediators}
\label{subsec:CEvNS_BSM}

NGIs constitute a useful model-independent probe that can accommodate several  attractive BSM scenarios. By restricting ourselves to low-energy  neutral-current interactions (below the electroweak symmetry breaking scale), in this work we consider general new physics interactions arising from the Lagrangian~\cite{Lindner:2016wff, AristizabalSierra:2018eqm}
\begin{equation}
\mathscr{L}_{NGI}=\frac{G_F}{\sqrt{2}}\sum_{\substack{X=S, P, V, A, T\\f=u,d\\ \alpha=e,\mu,\tau}}C_{\alpha,\alpha}^{f,P}\left[\bar{\nu}_\alpha\Gamma^X L \nu_\alpha\right]\left[\bar{f}\Gamma_X P f\right] \, .
 \label{equn:CEvNS_BSM_Effective_Lagrangian}
 \end{equation}
Therefore, in what follows all possible Lorentz-invariant structures are taken into account, i.e.  $\Gamma_X= \{\mathbb{1}, i\gamma_5, \gamma_\mu, \gamma_\mu\gamma_5, \sigma_{\mu\nu} \}$ (with $\sigma_{\mu\nu}=\frac{i}{2}[\gamma_\mu,\gamma_\nu])$, corresponding to  $ X= \{S, P, V, A, T\}$ interactions, respectively.
The dimensionless coefficients $C_{\alpha,\alpha}^{f,P}$ measure the relative strength of the new physics interaction  $X$ and are of the order of $(\sqrt{2}/G_F) (g_X^2/(q^2+m_X^2))$ with $m_X$ and $g_X$ being the mass of the exchanged light mediator and the corresponding coupling, respectively. Throughout this work, we define the coupling $g_X = \sqrt{g_{\nu X} g_{f X}}$; $f = \{ u, d\}$ for CE$\nu$NS and $f = e$ for E$\nu$ES.

We proceed by relying on previous analyses~\cite{Dienes:2013xya,Cirelli:2013ufw}  {that}, in the limit of vanishing momentum transfer, argued that the nucleonic matrix element of the quark current is proportional to that of the corresponding nucleon current  $\langle N_f| \bar{q} \Gamma^X q|N_i\rangle \equiv \mathcal{F}^X(q)\langle N_f| \bar{N} \Gamma^X N|N_i\rangle $, where $\mathcal{F}^X(q)$ is the form factor calculated within the framework of  {nonperturbative} low-energy QCD.
The individual interactions and their contribution to the \cevns differential cross section are listed in Table~\ref{table:CEvNS_BSM_Lagrangian_with_xsec}. 
For the different interactions $X$, the relevant effective charges $Q_X$ entering the respective cross sections are given by~\cite{Cirelli:2013ufw,Cerdeno:2016sfi,Demirci:2021zci,Barranco:2011wx} 
\begin{gather}
\begin{aligned}
Q_s=& g_{\nu S}\left(Z\sum_q g_{qS}\frac{m_p}{m_q}f^p_{q}+N\sum_q g_{qS}\frac{m_n}{m_q}f^n_{q}\right)F_W(q^2) \, ,\\
Q_P=& g_{\nu P}\left(Z\sum_q g_{qP}\frac{m_p}{m_q}h^p_{q}+N\sum_q g_{qP}\frac{m_n}{m_q}h^n_{q}\right)F_W(q^2) \, ,\\
Q_{V}=& g_{\nu V}\left[\left(2g_{uV}+g_{dV}\right)Z+\left(g_{uV}+ 2g_{dV} \right)N\right]F_W(q^2) \, ,\\
Q_A=& g_{\nu A}\left(Z S_p\sum_q g_{qA}\Delta^p_{q}+N S_n\sum_q g_{qA}\Delta^n_{q}\right)F_W(q^2) \, ,\\
Q_T=& g_{\nu T}\left(Z\sum_q g_{qT}\delta^p_{q}+N\sum_q g_{qT}\delta^n_{q}\right)F_W(q^2) \, .
\end{aligned}
\label{eq:BSM_charges_cevns}
\end{gather}
In the latter expressions, the hadronic structure parameters
for the scalar case ($S$) are $f_u^p=0.0208$, $f_u^n=0.0189$, $f_d^p=0.0411$,  {and} $f_d^n=0.0451$~\cite{AristizabalSierra:2019zmy},
for the pseudoscalar case ($P$) are $h^p_u=h^n_u=1.65$  {and} $h^p_d=h^n_d=0.375$~\cite{10.21468/SciPostPhys.10.3.072},
and for the tensor case ($T$) are $\delta^p_u=\delta^n_d=0.54$  {and} $\delta^p_d=\delta^n_u=-0.23$ \cite{AristizabalSierra:2019zmy}. 
Finally, for the case of axial vector ($A$) interactions the hadronic parameters read $\Delta^p_u=\Delta^n_u=0.842$  {and} $\Delta^p_d=\Delta^n_d=-0.427$ \cite{PhysRevD.77.065026},
while the spin expectation values $(S_p\text{ and }S_n)$ are nuclear model and isotope dependent. In the present study we consider the following spin expectation values for protons $(\mathrm{^{129}Xe}:S_p=0.010,~ \mathrm{^{131}Xe}:S_p=-0.009)$ and neutrons $(\mathrm{^{129}Xe}:S_n= 0.329,~ \mathrm{^{131}Xe}:S_n=-0.272)$, extracted from shell model nuclear structure calculations in Ref.~\cite{Hoferichter:2020osn}\footnote{Even-even Xenon isotopes with $0^+$ ground state do not contribute to axial vector interactions because of angular momentum conservation.}.

\begin{table}[t]
\begin{adjustbox}{width=\columnwidth,center}
\centering
\begin{tabular}{|c|c|c|}
\hline
Mediator & \hspace{0.5mm}$\mathscr{L}_X$ \hspace{0.5mm}& \hspace{0.05mm}Cross  {section}\hspace{0.05mm} \\
\hline
Scalar & $\left[\left(g_{\nu S}\bar{\nu}_R\nu_L+\text{ {H.c.}}\right)+\sum_{q=\{u,d\}}g_{qS}\bar{q}q\right]S+\frac{1}{2}m_S^2S^2$ & $\frac{m_N^2 E_{nr} Q_S^2}{4\pi E_\nu^2\left(q^2+m_S^2\right)^2}$  \\
\hline
Pseudoscalar & $\left[\left(g_{\nu P}\bar{\nu}_R\gamma_5\nu_L+\text{ {H.c.}}\right)-i\sum_{q=\{u,d\}}g_{qP}\bar{q}\gamma_5q\right]P+\frac{1}{2}m_P^2P^2$ & $\frac{m_N E_{nr}^2 Q_P^2}{8\pi E_\nu^2\left(q^2+m_P^2\right)^2}$ \\
\hline
Vector & $\left[g_{\nu V}\bar{\nu}_L\gamma_\mu\nu_L+\sum_{q=\{u,d\}}g_{qV}\bar{q}\gamma_\mu q\right]V^\mu+\frac{1}{2}m_V^2V^\mu V_\mu$ & $\left(1+\frac{Q_{V}}{\sqrt{2}G_FQ_V^{SM}\left(q^2+m_{V}^2\right)}\right)^2\left[\frac{d\sigma}{dE_{nr}}\right]_{SM}^{\nu N}$ \\
\hline
Axial vector & $\left[g_{\nu A}\bar{\nu}_L\gamma_\mu\gamma_5\nu_L-\sum_{q=\{u,d\}}g_{qA}\bar{q}\gamma_\mu\gamma_5 q\right]A^\mu+\frac{1}{2}m_A^2A^\mu A_\mu$ & $\frac{m_N Q_A^2(2E_\nu^2+m_N E_{nr})}{4\pi E_\nu^2\left(q^2+m_{A}^2\right)^2}$ \\
\hline
Tensor & \hspace{0.5mm}$\left[g_{\nu T}\bar{\nu}_R\sigma_{\rho\delta}\nu_L-\sum_{q=\{u,d\}}g_{qT}\bar{q}\sigma_{\rho\delta}q\right]T^{\rho\delta}+\frac{1}{2}m_T^2T^{\rho\delta}T_{\rho\delta}$\hspace{0.5mm} & $\frac{m_N Q_T^2(4E_\nu^2-m_N E_{nr})}{2\pi E_\nu^2\left(q^2+m_{T}^2\right)^2}$  \\
\hline
\end{tabular}
\end{adjustbox}
\caption{Novel interactions $X = \{S,P,V,A,T \}$ and corresponding differential CE$\nu$NS cross sections considered in the present work.  {Because of} interference, the $V$ interaction is the only case that includes the SM contribution, while the $S, P, A, T$ cases acquire contributions from new physics only, as shown in the cross sections (see the text for more details).}
\label{table:CEvNS_BSM_Lagrangian_with_xsec}
\end{table}

A few comments are in order. First, from Table~\ref{table:CEvNS_BSM_Lagrangian_with_xsec} it becomes evident that for the case of $S,P,A,T$ interactions there is  {an absence of interference} with the SM \cevns cross section, in contrast to the vector mediator case where the new physics contribution yields interference terms. For the sake of clarity we should stress that, in principle, the axial vector interaction adds incoherently to the SM \cevns cross section, however the SM axial contribution is significantly suppressed with respect to the  vector one and therefore neglected (see Ref.~\cite{Papoulias:2017qdn}). Moreover, as discussed in Ref.~\cite{AristizabalSierra:2018eqm}, the existence of axial quark terms in the interaction Lagrangian~(\ref{equn:CEvNS_BSM_Effective_Lagrangian}) will lead to pseudoscalar-scalar neutrino-quark couplings, the study of which will be relaxed as it goes well beyond the scope of our work. Finally, the $P$ and $A$ terms are spin-dependent and therefore suppressed with respect to  $S,V,T$ terms\footnote{In Ref.~\cite{Cerdeno:2016sfi} the $P$ contribution to \cevns was reported to be vanishing based on the fact that the corresponding nucleon matrix element is vanishing~\cite{Cheng:1988im,Cheng:2012qr}.}. For the latter issue, we will rely on the assumption that the new physics interaction is controlled by the strength of the neutrino coupling.

Focusing on light vector mediators only, in this work we will consider three different models, determined by the charges of leptons $Q'_\ell$ and quarks $Q'_q$ under the extra gauge symmetry ${U}(1)'$~\cite{Flores:2020lji}. First, we consider a generic  model where the new vector mediator $V$ couples universally to all SM fermions, such that $Q'_\ell=Q'_q=1$, corresponding to an effective vector charge in Eq.(\ref{eq:BSM_charges_cevns}) with $g_{q V}=g_{\nu V}$.  {Second}, we focus on the $U_{B-L}$ extension of the SM, in which the anomaly cancellation conditions $Q'_\ell=-1, ~Q'_q = 1/3$ imply an effective vector charge with $g_{q V} = -g_{\nu V}/3$.  Finally, we consider the gauged $U_{L_\mu-L_\tau}$ symmetry, where the new vector mediator boson couples directly only to muons and tauons  with  {the} absence of tree-level couplings to  quarks. In this model, contributions to \cevns are possible at the  {one-loop} level,  leading to an ``effective'' kinetic mixing between the new mediator and the SM photon~\cite{Altmannshofer:2019zhy}, and the corresponding cross section reads~\cite{Cadeddu:2020nbr, Corona:2022wlb} 
\begin{align}
\label{equn:CEvNS_Lmu-Ltau_xsec}
\begin{split}
\left[\frac{d\sigma}{d E_{nr}}\right]_{L_\mu-L_\tau}^{\nu N}=&\left(1 \pm \frac{\alpha_{em}g_{\nu V}g_{qV}\log\left(\frac{m_\tau^2}{m_\mu^2}\right)Z F_W(q^2) }{3\sqrt{2}\pi G_FQ_V^{SM}\left(q^2+m_{V}^2\right)}\right)^2\left[\frac{d\sigma}{d E_{nr}}\right]_{SM}^{\nu N} \, ,
\end{split}
\end{align}
where $\alpha_{em}$ is  {the} fine structure constant,  {and} $m_\mu$ and $m_\tau$ denote the muon and tau masses, respectively\footnote{For a study focusing on models with an extra $U_{B-2 L_\alpha-L_\beta}$ and $U_{B-3 L_\alpha}$ gauge symmetry, see Ref.~\cite{delaVega:2021wpx}.}, while the plus (minus) sign accounts for $\nu_\mu$-nucleus ($\nu_\tau$-nucleus) scattering.

\section{Elastic Neutrino-Electron Scattering}
\label{sec:EveS}
Proceeding in an analogous way as in the case of CE$\nu$NS, in this section we discuss the formalism describing \eves within  and beyond the SM. 
\subsection{E$\nu$ES through SM interaction Channel}
\label{subsec:EveS_SM}
E$\nu$ES is a well-understood weak  {flavored} process where ( {anti})neutrinos of  {flavor} $\alpha = \{e, \mu, \tau\}$ interact with electrons via elastic scattering at low and intermediate energies, receiving contributions from both  {neutral and charged currents} for $\alpha=e$, and  {neutral current} only for $\alpha = \{\mu, \tau\}$. Within the framework of the SM, the Lagrangian density corresponding to the process $\overset{\scriptscriptstyle(-)}{\nu_e}+e^-\rightarrow \overset{\scriptscriptstyle(-)}{\nu_e}+e^-$, reads\footnote{For the case of $\nu_{\mu,\tau}$--$e^{-}$ \eves the Lagrangian involves only neutral-current terms.}~\cite{Giunti:2007ry}
\begin{align}
\label{equn:EveS_SM_Lagrangian_for_nu-e}
\begin{split}
 \mathscr{L}_{SM}(\overset{\scriptscriptstyle(-)}{\nu_e}+e^-\rightarrow \overset{\scriptscriptstyle(-)}{\nu_e}+e^-)=&-\frac{G_F}{\sqrt{2}}\{\left[\bar{\nu}_e\gamma^\rho \left(1-\gamma_5\right)e\right]\left[\bar{e}\gamma_\rho \left(1-\gamma_5\right) \nu_e\right]\\&+\left[\bar{\nu}_e\gamma^\rho \left(1-\gamma_5\right)\nu_e\right]\left[\bar{e}\gamma_\rho \left(g_V-g_A\gamma_5\right) e\right]\} \, ,
\end{split}
\end{align}
while at the tree level the corresponding differential cross section with respect to the electron recoil energy $E_{er}$, is given as~\cite{Kayser:1979mj}
\begin{align}
\label{equn:EveS_SM_xsec_for_nu-alpha}
\begin{split}
\left[\frac{d\sigma_{\nu_\alpha}}{d E_{er}}\right]_{SM}^{\nu e}=&\frac{G_F^2m_e}{2\pi}[(g_V + g_A)^2 +(g_V - g_A)^2\left(1-\frac{E_{er}}{E_\nu}\right)^2 -(g_V^2-g_A^2)\frac{m_e E_{er}}{E_\nu^2}] \, .
\end{split}
\end{align}
In Eq.(\ref{equn:EveS_SM_xsec_for_nu-alpha}), $g_V$ and $g_A$ are vector and axial vector couplings respectively, and take the form
\begin{equation}
 \label{table:EveS_SM_couplings}
 g_V=-\frac{1}{2}+2\sin^2\theta_W + \delta_{\alpha e}, \qquad  g_A=-\frac{1}{2}+ \delta_{\alpha e} \, ,
\end{equation}
where the $\delta_{\alpha e}$ term is the Kronecker delta, which becomes nonzero only for $\nu_e$--$e^{-}$ interactions. For the case of antineutrino scattering, the \eves cross section is given by Eq.(\ref{equn:EveS_SM_xsec_for_nu-alpha}) with the substitution $g_A \to - g_A$.

\subsection{E$\nu$ES through light novel mediators}
\label{subsec:EveS_BSM}

Similar to the CE$\nu$NS case, here also we consider all possible \eves contributions arising from the  Lorentz-invariant forms $X=\{S,P,V,A,T\}$.  For light vector and axial vector novel mediators, the total differential cross sections can be achieved by replacing $g_V$ and $g_A$ from the SM cross section as~\cite{Lindner:2018kjo,Ballett:2019xoj} 
\begin{equation}
\label{equn:EveS_BSM_xsec_V,A}
g'_{V/A}=g_{V/A}+\frac{g_{\nu V/A}\cdot g_{e V/A}}{4\sqrt{2}G_F(2m_e E_{er} + m_{V/A}^2)} \, .
\end{equation}
It is interesting to notice that in the $U_{B-L}$ gauge extension the cross section is also given by Eq.(\ref{equn:EveS_BSM_xsec_V,A}) since $Q'_\ell=Q'_\nu=-1$, while in the universal case it holds that $Q'_\ell=Q'_\nu=1$, hence both cases leading to $g_{\nu V/A} = g_{e V/A}$ in Eq.(\ref{equn:EveS_BSM_xsec_V,A}).  On the other hand, under  $U_{L_\mu - L_\tau}$ symmetry, the new light vector mediator contributes at  {one-loop} level for $\sigma_{\nu_\mu - e}$ and $\sigma_{\nu_\tau - e}$ only, while $\sigma_{\nu_e - e}$ is vanishing\footnote{ {Nonzero} couplings contribute to $\sigma_{\nu_e - e}$ at  {two-loop} level through $Z_0-V$ mixing, which are neglected in this work.}. The relevant couplings read~\cite{Altmannshofer:2019zhy}  
\begin{equation}
g'_V=  g_V \pm \frac{\alpha_{em}}{3 \sqrt{2} \pi G_F} \log{\left(\frac{m^2_\tau}{m_\mu^2} \right)} \frac{g_{\nu V}\cdot g_{e V}}{(2m_e E_{er} + m_{V}^2)} \, ,
\label{equn:EveS_BSM_xsec_LVM_LmLt}
\end{equation}
where, unlike the \cevns case, the plus (minus) sign corresponds to $\sigma_{\nu_\tau - e}$ $(\sigma_{\nu_\mu - e})$ scattering.
The remaining $S$, $P$, and $T$ cross section contributions add incoherently to the SM cross section and have been previously written as~\cite{Khan:2020vaf, Link:2019pbm}
\begin{equation}
\left[\frac{d\sigma_{\nu_\alpha}}{d E_{er}}\right]_{S}^{\nu e}=\left[\frac{g^2_{\nu S}\cdot g^2_{eS}}{4\pi(2m_e E_{er} + m_{S}^2)^2}\right]\frac{m_e^2 E_{er}}{E_\nu^2}\, ,
\end{equation}
\begin{equation}
\left[\frac{d\sigma_{\nu_\alpha}}{d E_{er}}\right]_{P}^{\nu e}= \left[  \frac{g^2_{\nu P}\cdot g^2_{eP}}{8\pi(2m_e E_{er} + m_{P}^2)^2} \right]\frac{m_e E_{er}^2}{E_\nu^2}\, ,
\end{equation}
\begin{equation}
\label{equn:EvES_Tensor}
\left[\frac{d\sigma_{\nu_\alpha}}{d E_{er}}\right]_{T}^{\nu e}=\frac{m_e\cdot g^2_{\nu T}\cdot g^2_{eT}}{2\pi(2m_e E_{er} + m_{T}^2)^2}\cdot\left[1+2\left(1-\frac{E_{er}}{E_\nu}\right)+\left(1-\frac{E_{er}}{E_\nu}\right)^2-\frac{m_e E_{er}}{E_\nu^2}\right]\, .
\end{equation}

\section{Results}
\label{sec:results}
\subsection{\cevns and \eves events at direct detection dark matter detectors}
\label{subsec:nu_flux}

In our analysis we consider astrophysical neutrinos coming from  the  {Sun}~\cite{Haxton:2012wfz}, the atmosphere~\cite{Battistoni:2005pd}, and diffuse supernovae (DSN)~\cite{Beacom:2010kk}. In our calculations we have neglected the neutrino contributions coming from Earth, known as  {geoneutrinos}, as their induced events are expected to be overshadowed by several orders of magnitude with respect to solar neutrinos~\cite{Monroe:2007xp,Gelmini:2018gqa,Kosmas:2021zve}.
For the  {normalization} of the different neutrino fluxes we use the recommended conventions reported in Ref.~\cite{Baxter:2021pqo} which are listed in Table~\ref{table:neutrino_flux_normalisation}.
\begin{table}[t]
\centering
\begin{tabular}{|c|c|c|}
\hline
Component & $E^{max}_\nu$~(MeV) & Flux~($\mathrm{cm^{-2}s^{-1}}$)\\
\hline
$pp$ & $0.42341$ & $5.98\times 10^{10}$\\

$pep$ & $1.44$ & $1.44\times 10^{8}$\\
$^7\mathrm{Be_{\text{High}}}$ & $0.8613$ &  $4.35\times 10^{9}$\\
$^7\mathrm{Be_{\text{Low}}}$ & $0.3843$ &  $4.84\times 10^{8}$\\
$^8$B & $16.36$ &  $5.25\times 10^{6}$\\
$hep$ & $18.784$ &  $7.98\times 10^{3}$\\
$^{13}$N & $1.199$ &  $2.78\times 10^{8}$\\
$^{15}$O & $1.732$ &  $2.05\times 10^{8}$\\
$^{17}$F & $1.74$ &  $5.29\times 10^{6}$\\
 {Atm.} & $981.75$ &  $10.5$\\
DSN & $91.201$ &  $86$\\
\hline
\end{tabular}
\caption{Neutrino end point energy and flux normalization for the different astrophysical neutrino sources. The flux normalizations are taken from Ref.~\cite{Baxter:2021pqo}.}
\label{table:neutrino_flux_normalisation}
\end{table}

For the interaction channel $X$, the differential event rate of CE$\nu$NS at a given detector follows from the convolution of the differential cross section with the neutrino energy distribution $d \Phi/dE_\nu$, as\footnote{An ideal detector with perfect efficiency and resolution power is assumed.}~\cite{Papoulias:2018uzy}
\begin{equation}
\label{equn:CEvNS_Differential_Event_Rate}
\left[\frac{dR}{d E_{nr}}\right]_X^{\nu N} =t_{run}N_{target} \mathcal{A}(E_{nr}) \sum_i \int_{E_\nu^{min}}^{E_\nu^{max}}dE_{\nu} \, \frac{d\Phi_i(E_\nu)}{dE_\nu}\left[\frac{d\sigma}{d E_{nr}}(E_\nu ,E_{nr})\right]_X^{\nu N} \, ,
\end{equation}
where $\mathcal{A}(E_{nr})$ is the detection efficiency, $t_{run}$ denotes the exposure time and $N_{target}$ represents the number of target nuclei. In the latter expression, the index $i$ runs over all the  neutrino sources with energy distribution $d\Phi_i/dE_\nu$ and $E_\nu^{max}$ denotes the maximum neutrino energy of the $i$th source (see Table \ref{table:neutrino_flux_normalisation}). Finally, the minimum neutrino energy $E_\nu^{min}$ required to generate a nuclear recoil with energy $E_{nr}$ is trivially obtained from the kinematics of the process and reads
 \begin{equation}
 \label{equn:Min_Neutrino_Energy}
 E_\nu^{min}= \frac{1}{2}\left[E_{nr}+\sqrt{E_{nr}^2+2m_N E_{nr}}\right] 
 \approx\sqrt{\frac{m_N E_{nr}}{2}} \, .
 \end{equation}
In our analysis we also consider corrections from detector-specific quantities. In particular, for the  {ionization} xenon detectors considered here, a significant amount of nuclear recoil energy $E_{nr}$ is lost into heat and other dissipative energies, so that the actual energy measured by the detector is an \emph{electron equivalent} energy $E_{er}$~\cite{Lewin:1995rx}.  In our calculation this effect is taken into account through the quenching factor $\mathcal{Q}_f(E_{nr})$, calculated on the basis of theoretical predictions within Lindhard theory~\cite{Lindhard:1963}
 \begin{equation}
 \label{equn:Quenching_Factor}
\frac{E_{er}}{E_{nr}}=\mathcal{Q}_f{(E_{nr})}=\frac{kg(\gamma)}{1+kg(\gamma)} \, ,
 \end{equation}
where $g(\gamma)=3\gamma^{0.15}+0.7\gamma^{0.6}+\gamma$, with $\gamma=11.5\cdot Z^{-\frac{7}{3}}E_{nr} (\mathrm{keV_{nr}})$ and $k=0.133\cdot Z^{\frac{2}{3}}A^{-\frac{1}{2}}$. Let us stress that for the case of germanium, low-energy corrections to the quenching factor can been accounted for through the adiabatic correction of Lindhard theory, introduced in Ref.~\cite{Scholz:2016qos}\footnote{\scriptsize{Because of the lack of data for the xenon isotope we are interested in this work, we have verified that our results remain unaffected even when considering $\pm 1\sigma$ deviations from the adiabatic parameter corresponding to germanium (see Table I of Ref.~\cite{Scholz:2016qos}). Improved quenching factors at sub-keV energies are comprehensively discussed in Ref.~\cite{Sarkis:2020soy}.}}. After incorporating the quenching factor corrections, the number of events in the $j$th bin is written as
 \begin{equation}
 \label{equn:binned_number_of_quenched_events}
\left[ R_j \right]_X^{\nu N} =\int_{E_{er}^j}^{E_{er}^{j+1}} dE_{er} \, \left[\frac{dR}{d E_{nr}}\right]_X^{\nu N} \frac{1}{\mathcal{Q}_f}\left(1-\frac{E_{er}}{{\mathcal{Q}_f}}\frac{d \mathcal{Q}_f}{d E_{er}}\right)  \, ,
 \end{equation}

At this point we turn our attention to E$\nu$ES. For the case of solar neutrinos the differential number of events takes into account the effect of neutrino oscillations in propagation and is given according to the expression~\cite{AristizabalSierra:2017joc}
\begin{equation}
\left[\frac{dR}{dE_{er}}\right]_X^{\nu e} =t_{run}N_{target}\mathcal{A}(E_{er})\sum_{i = \mathrm{solar}}\int_{E_\nu^{min}}^{E_\nu^{max}}dE_{\nu}\frac{d\Phi^{\nu_e}_i(E_\nu)}{dE_\nu}\left(P_{ee}{\left[\frac{d\sigma_{\nu_{e}}}{d E_{er}}\right]_X^{\nu e}}  + \overline{P_{e f}} {\left[\frac{d\sigma_{\nu_{f}}}{d E_{er}}\right]_X^{\nu e}} \right) \,,
\label{equn:EvES_Differential_Event_Rate}
\end{equation}
with $f=\mu, \tau$ and the minimum neutrino energy being
\begin{equation}
 \label{equn:Min_Neutrino_Energy2}
 E_\nu^{min}=\frac{1}{2}\left[E_{er}+\sqrt{E_{er}^2+2m_e E_{er}}\right] \, .
 \end{equation}
Since the \eves cross section is not  {flavor} blind as the \cevns one, Eq.(\ref{equn:EvES_Differential_Event_Rate}) incorporates neutrino oscillations by weighting the  {flavored} cross section with the corresponding oscillation probability. For our purposes it is sufficient to consider the  averaged oscillation probability $P_{ee}(E_\nu)$ in the  {two-flavor} approximation, which we take from~\cite{Escrihuela:2009up}. Notice that in the second term of Eq.(\ref{equn:EvES_Differential_Event_Rate}) the oscillation factors read $\overline{P_{e \mu}} \equiv (1-P_{ee}) \cos^2 \theta_{23}$ and $\overline{P_{e \tau}} \equiv (1-P_{ee}) \sin^2 \theta_{23}$ for incoming $\nu_\mu$ and $\nu_\tau$ neutrinos, respectively\footnote{The value of the atmospheric mixing angle ($\theta_{23}$) is taken from the best fit of the 2020 Valencia global-fit~\cite{deSalas:2020pgw}.}.
Let us note that for the case of atmospheric and DSN neutrinos, oscillation effects are neglected. This is a reasonable approximation given the fact that the expected \eves event rates from atmospheric and DSN neutrinos are suppressed by several orders of magnitude compared to the dominant solar ones (see the discussion below). Therefore in our statistical analysis this approximation is not expected to have any quantitative effect. The differential number of \eves events relevant to atmospheric and DSN neutrinos is then calculated as
\begin{equation}
\left[\frac{dR}{d E_{er}}\right]_X^{\nu e}  =t_{run}N_{target} \mathcal{A}(E_{er})\sum_{i = \mathrm{atm, DSN}}\int_{E_\nu^{min}}^{E_\nu^{max}}dE_{\nu}\frac{d\Phi_i^{\nu_\alpha}(E_\nu)}{dE_\nu} \left[\frac{d\sigma_{\nu_\alpha}}{d E_{er}}\right]_X^{\nu e} \, , \quad \nu_\alpha = \{\nu_e, \bar{\nu}_e, \nu_{\mu, \tau}, \bar{\nu}_{\mu, \tau}\} \,.
\end{equation}
\begin{table}[t]
\centering
\begin{tabular}{|c|c||c|c||c|c||c|c|}
\hline
\hspace{0.05mm} State \hspace{0.05mm}& \hspace{0.1mm}  {SP} (eV)\hspace{0.1mm} & \hspace{0.05mm} State \hspace{0.05mm} & \hspace{0.1mm}  {SP} (eV) \hspace{0.1mm} & \hspace{0.05mm} State \hspace{0.05mm} & \hspace{0.1mm}  {SP} (eV) \hspace{0.1mm} & \hspace{0.05mm} State \hspace{0.05mm} & \hspace{0.1mm}  {SP} (eV)\hspace{0.1mm} \\
\hline
$^1s_{\frac{1}{2}}$ & $34759.3$ & $^3p_{\frac{3}{2}}$ & $1024.8$ & $^4p_{\frac{3}{2}}$ & $708.1$ & $^5p_{\frac{1}{2}}$ & $13.4$\\
\hline
$^2s_{\frac{1}{2}}$ & $5509.8$ & $^3p_{\frac{1}{2}}$ & $961.2$ & $^4p_{\frac{1}{2}}$ & $162.8$ & $^5p_{\frac{3}{2}}$ & $12.0$\\
\hline
$^2p_{\frac{3}{2}}$ & $5161.5$ & $^3d_{\frac{5}{2}}$ & $708.1$ & $^4d_{\frac{5}{2}}$ & $73.8$ &  & \\
\hline
$^2p_{\frac{1}{2}}$ & $4835.6$ & $^3d_{\frac{3}{2}}$ & $694.9$ & $^4d_{\frac{3}{2}}$ & $71.7$ &  & \\
\hline
$^3s_{\frac{1}{2}}$ & $1170.5$ & $^4s_{\frac{1}{2}}$ & $229.4$ & $^5s_{\frac{1}{2}}$ & $27.5$ &  & \\
\hline
\end{tabular}
\caption{ {Single-particle} ({SP}) energies for the  {xenon} atom derived from  Hartree-Fock calculations in Ref.~\cite{Chen:2016eab}.}
\label{table:Xe_e_Binding_Energy}
\end{table}

Up to this point, for all \eves interaction channels within and beyond the SM, our discussion applies to the case where neutrinos scatter off free electrons. However, the target electrons are not free, but rather bounded inside the atom. Therefore, in order to perform  realistic simulations of the expected number of events at a given detector, atomic binding effects should be also considered.  To account for binding effects in our analysis, the free \eves cross section $(d\sigma_{\nu_\alpha}/d E_{er})_\text{free}$ defined in Eqs.\eqref{equn:EveS_SM_xsec_for_nu-alpha}-\eqref{equn:EvES_Tensor} is weighted by a series of step functions introduced in Ref.~\cite{Chen:2016eab} as follows
 \begin{equation}
 \label{equn:step_fn}
\left[\frac{d\sigma_{ \nu_\alpha}}{dE_{er}}\right]_X^{\nu e}=\frac{1}{Z}\sum_{i=1}^{Z} \Theta (E_{er}-B_i)\, \left(\left[\frac{d\sigma_{\nu_\alpha}}{d E_{er}}\right]_X^{\nu e}\right)_\text{free} \, .
 \end{equation}
 Here, $\Theta(x)$ is the Heaviside step function, while the quantity $\sum_{i=1}^Z \Theta (E_{er}-B_i)$ quantifies the number of electrons that can be  {ionized} by the recoil energy $E_{er}$ and  $B_i$  {represents} the binding energy of the  {$i$th} electron in the atom. The  {single-particle} atomic level binding energy of electrons in  {a }$^{131}$Xe atom is given in Table~\ref{table:Xe_e_Binding_Energy}~\cite{Chen:2016eab}.

\begin{figure}[!htbp]
\begin{center}
\begin{tabular}{lr}
 \vspace{-1.5cm}
\hspace{-1.5cm}
\\ \begin{subfigure}[b]{0.99\textwidth}
\includegraphics[width=0.49\textwidth]{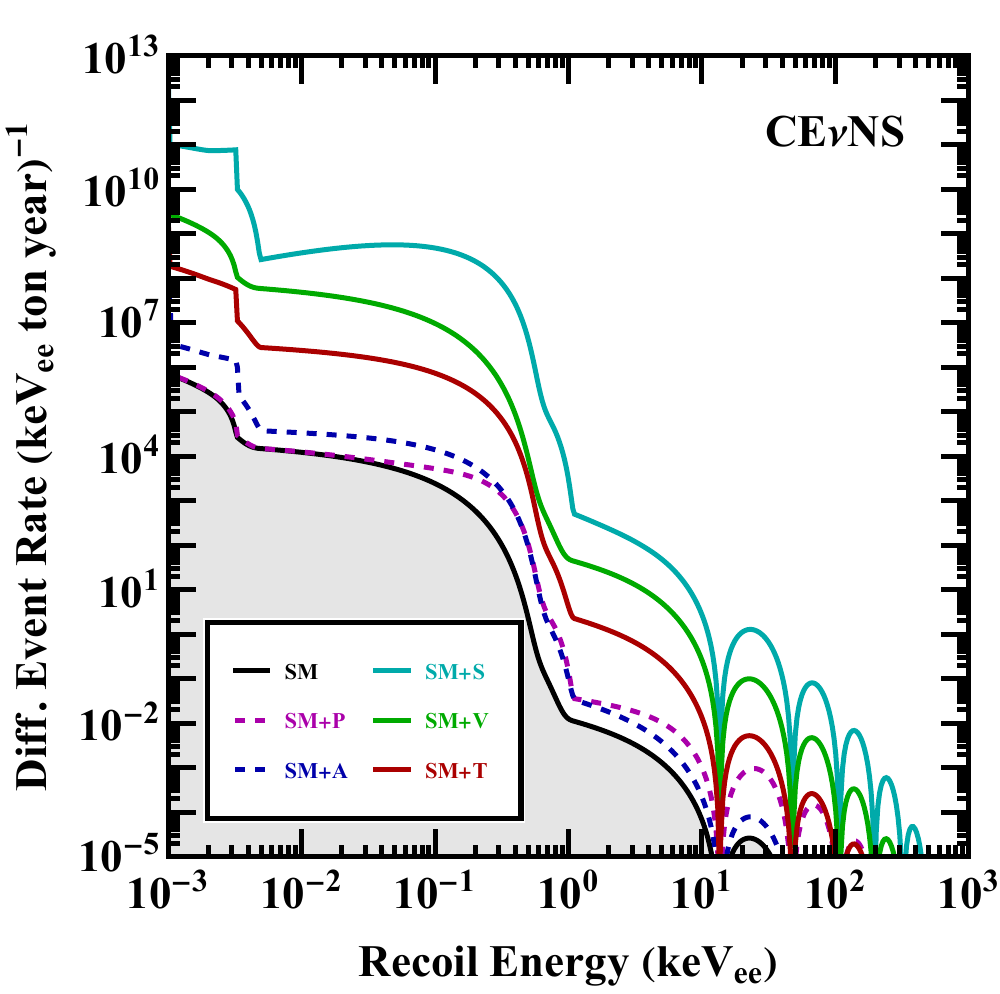}
\includegraphics[width=0.49\textwidth]{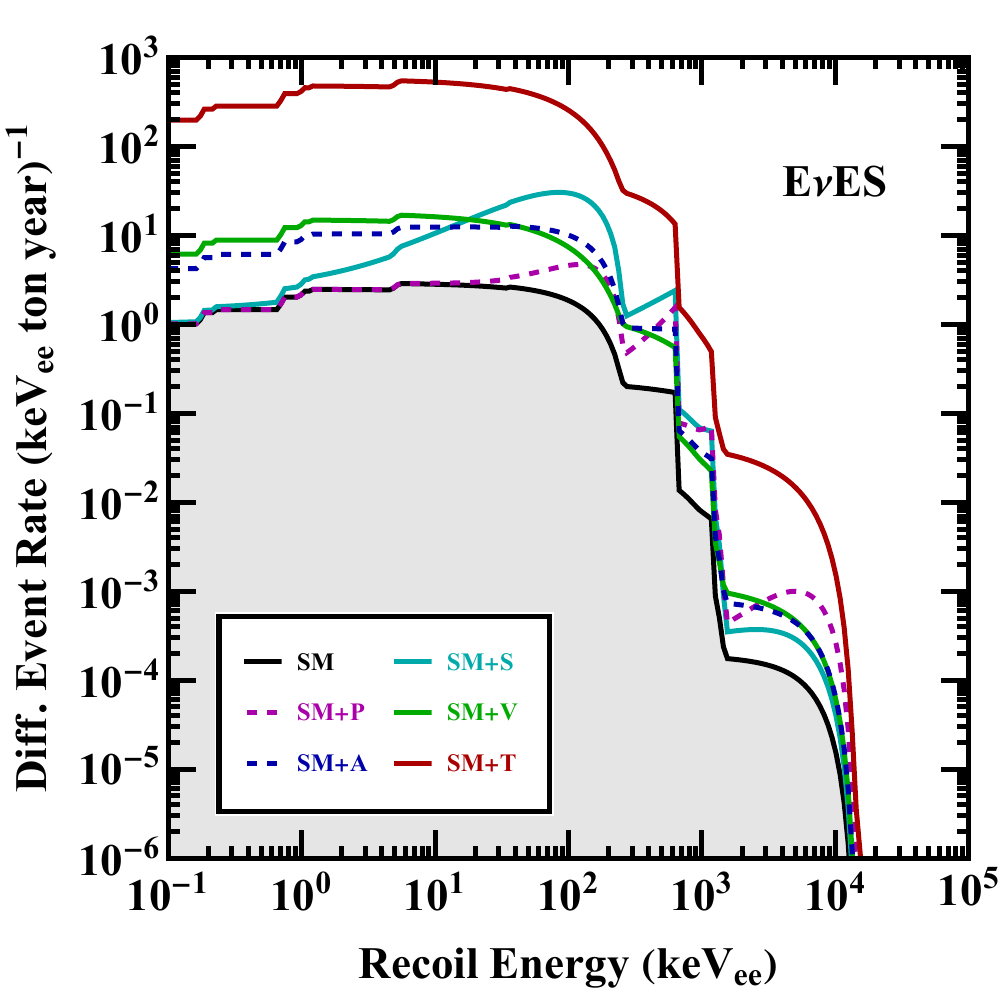}
\captionsetup{justification=centering}
\caption[]{Differential event spectra for CE$\nu$NS (left) and E$\nu$ES (right) processes.}
\vspace{-0.01cm}
\end{subfigure}\\
\vspace{-0.5cm} 
\begin{subfigure}[b]{0.99\textwidth}
\includegraphics[width=0.49\textwidth]{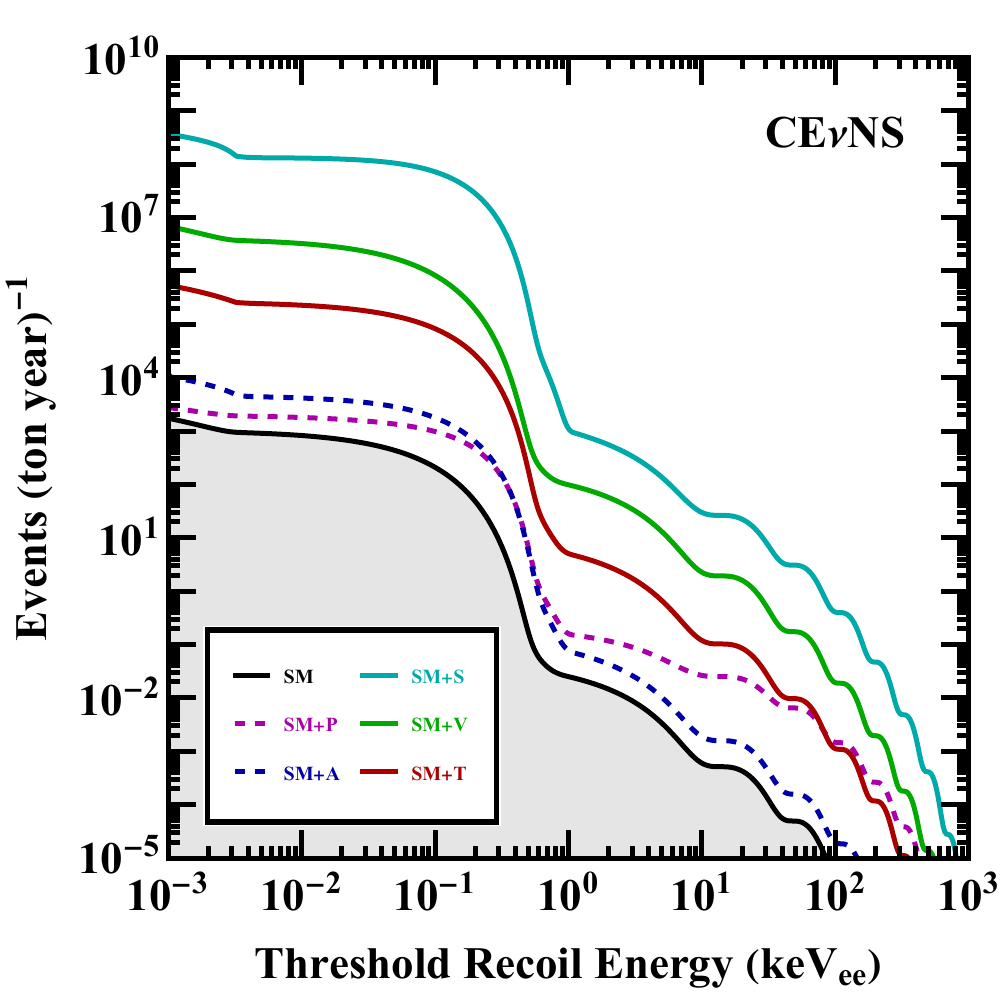}
\includegraphics[width=0.49\textwidth]{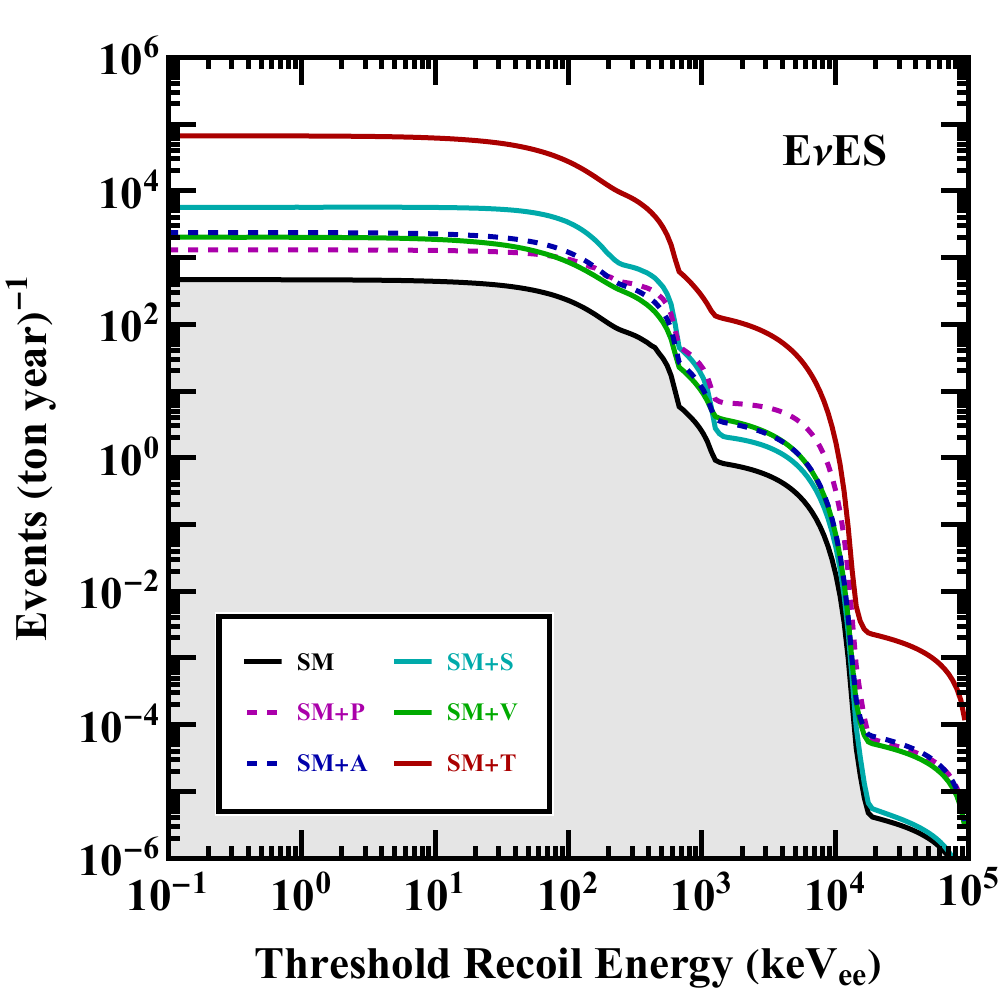}
\captionsetup{justification=centering}
\caption[]{Integrated event spectra for CE$\nu$NS (left) and E$\nu$ES (right) processes.}
        \label{fig:bp3}
\end{subfigure}
\end{tabular}
\end{center}
\caption{Differential (a) and integrated (b) event spectra expected at a xenon target for \cevns and \eves processes as a function of the recoil energy (threshold recoil energy) measured in $\mathrm{keV_{ee}}$ units. The contributions from the different new physics interactions $X$ are calculated assuming the benchmark values $g^2_X=10^{-4}$ and $m_X=1$~GeV (for details see the text).}
\label{fig:events}
\end{figure}
 
%

Assuming typical benchmark values $g_X^2=10^{-4}$ and $m_X=1$~GeV, differential and integrated \cevns (E$\nu$ES) event rates above threshold for a xenon target (assuming average isotopic abundance) are shown in the left (right) panel of Fig.~\ref{fig:events} as functions of the recoil energy for all possible interactions $X$. It is interesting to notice that, in general, the \cevns number of events are expected  to exceed the \eves rate by several orders of magnitude. However, for a realistic next  generation experiment operating with typical threshold energies i.e. $E_{er}>0.1~\mathrm{keV_{ee}}$, this is not strictly true. First, for $E_{er}>0.1~\mathrm{keV_{ee}}$ it becomes evident that CE$\nu$NS-induced events will be dominated by $^{8}$B neutrinos while the E$\nu$ES ones will be dominated by $pp$ neutrinos. Then, from a closer inspection of the graphs it can be deduced that \cevns and \eves are expected to generate a comparable number of events (at least) within the SM. This can be understood as follows. Although the \cevns cross section scales with $\sim N^2$ as a consequence of the coherency in neutrino-nucleus scattering, the E$\nu$ES-induced number of events is enhanced by an overall multiplicative factor $Z$ coming due to the number of electron targets, while the remaining difference, which is of the order of an overall factor $\sim N$, is compensated by the relative difference of $^{8}$B and $pp$ flux  {normalization}. Before closing this discussion, let us note the impact of nuclear (atomic) effects in the \cevns (E$\nu$ES) rates. In particular, as can be seen from  the \cevns spectral rates, the various dips  {occurring} for $E_{er}>25~\mathrm{keV_{ee}}$ are due to the minima of the nuclear form factor (see Fig.~3 of Ref.~\cite{Sahu:2020kwh}), while the slight suppression of the \eves rates occurring for recoil energies below $\sim 35~\mathrm{keV_{ee}}$ reflects the atomic binding effects (see Table~\ref{table:Xe_e_Binding_Energy}).

\subsection{Sensitivity on $S, P, V, A, T$ interactions}
\label{subsec:chi_squre}

At this point we are interested to explore the projected sensitivities on the ($m_X, g_X$) parameter space at a typical next generation experiment looking for direct detection of dark matter. For estimating the projected sensitivities of the next generation experiments to the various new interaction channels $X=\{S,P,V,A,T\}$, we perform a spectral fit in terms of the $\chi^2$ function~\cite{Feldman:1997qc}
\begin{equation}
\begin{aligned}
\chi^2(\xi, m_X,g_X)=2\sum_{i=1}^{n_\mathrm{bins}} \Biggl[  R_\mathrm{pred}^i(\xi, m_X,g_X) - R_\mathrm{exp}^i  & +R_\mathrm{exp}^i\ln \left(\frac{R_\mathrm{exp}^i}{ R_\mathrm{pred}^i(\xi, m_X,g_X)}\right) \Biggr]+\left(\frac{\xi}{\sigma_\text{sys}}\right)^2 \, .
\end{aligned}
\label{equn:chi_square}
\end{equation}
where $\xi$ denotes a nuisance parameter to account for the various systematic uncertainties.
The expected number of events in the $i$th bin is defined as $R_\mathrm{exp}^i= R_\text{SM}^i + R_\text{bg}^i$ with $R_\text{bg}^i$ being the number of background events, while the predicted number of events in the presence of the interaction $X$ is  $R_\mathrm{pred}^i= (1+\xi)\left[R_\text{X+SM}^i(m_X,g_X) + R_\text{bg}^i\right]$.  At this point we should stress that in our E$\nu$ES-based statistical analysis, the atmospheric and DSN fluxes are safely neglected. Indeed, this is a reasonable approximation given the expected SM event rates depicted in the right panel of Fig.~\ref{fig:events}. In the present work we will consider two representative case studies accounting for different detector-specific quantities, namely an ``optimistic" and a ``realistic" scenario. The two scenarios differ in the details of the detector properties and effects that are taken into account. One of our aims is to compare and contrast the results obtained from  {an} ideal vs realistic detector. The details of the two scenarios are as described below.  

\begin{itemize}

\item \textbf{Optimistic scenario:} A xenon detector with a $1~\mathrm{ton \cdot yr}$ exposure is assumed. This is, however, a conservative choice  {that} exceeds slightly the $0.65~\mathrm{ton \cdot yr}$ achieved by XENON1T~\cite{XENON:2020rca} and by a factor 20 less than the exposure goal of $20~\mathrm{ton \cdot yr}$ at XENONnT~\cite{XENON:2020kmp}. We stick to the latter choice to avoid potential overestimation of our projected sensitivities, since for this scenario the detector efficiency and energy resolution are not taken into consideration. For both \cevns and \eves analyses, we consider a flat background taken to be 10\% of the SM events as well as a systematic uncertainty $\sigma_\text{sys}=10\%$.  For our spectral fit we consider 100 log-spaced bins in the range $(0.1,100)~\mathrm{keV_{ee}}$ for the case of \cevns and in the range $(0.1,10^4)~\mathrm{keV_{ee}}$ for the case of E$\nu$ES. Let us clarify that our optimistic scenario assumes a threshold as low as $0.1~ \mathrm{keV_{ee}}$ and a perfect efficiency, which is motivated by the ``$S2$-only" analysis done in Ref.~\cite{Ni:2021mwa}. For the case of \eves we have also checked that our projected sensitivities remain the same if a narrower recoil window $(0.1,100)~\mathrm{keV_{ee}}$ is chosen.

\item \textbf{Realistic scenario:} This scenario will also serve as reference point to highlight and contrast the results expected from an ideal and a realistic detector. We consider the planned $20~\mathrm{ton \cdot yr}$ exposure foreseen at XENONnT, and we incorporate realistic backgrounds, efficiency and resolution effects. In particular, for our \cevns analysis  we consider the neutron background\footnote{For this particular case, due to lack of information we do not assign any uncertainty on the neutron background i.e., our predicted number of events in this case reads $R_\mathrm{pred}^i= (1+\xi) \left[R_\text{X+SM}^i(m_X,g_X) \right]+ R_\text{bg}^i$. We have checked, however, that our result remains  {essentially} unaltered if a background uncertainty as high as 10\% is assumed.} from Ref.~\cite{XENON:2020kmp} and the efficiency from Ref.~\cite{XENON:2018voc}.  For our \eves analysis, the background model $B_0$ (appropriately scaled to account for the $20~\mathrm{ton \cdot yr}$ exposure) and the detection efficiency, are both taken from Ref.~\cite{XENON:2020rca}. Moreover, our theoretical \eves event rates are eventually smeared\footnote{This requires another integration of Eq.(\ref{equn:EvES_Differential_Event_Rate}) over the true recoil energy $E_{er}$ taking the resolution function to be a  {normalized} Gaussian, see Ref.~\cite{Miranda:2020kwy}.} by taking into account the energy resolution power of XENON1T, that is $\sigma(E) = a \cdot \sqrt{E} + b  \cdot E$, with $E$ being the reconstructed recoil energy, $a=0.310~\sqrt{\mathrm{keV_{ee}}}$,  {and} $b = 0.0037~\mathrm{keV_{ee}}$~\cite{XENON:2020rca}. We consider 17 linearly spaced bins in the range $(4,52)~\mathrm{keV_{nr}}$ for our \cevns analysis and 30 linearly spaced bins in the range  $(0,30)~\mathrm{keV_{ee}}$ for our \eves analysis~\cite{XENON:2020kmp}. Finally, a 10\% systematic uncertainty is assigned for \cevns as in the optimistic scenario, while for the case of \eves a 3\% systematic uncertainty is taken to account for the detection uncertainty of XENON1T.
\end{itemize}

\begin{figure}[!htbp]
\begin{center}
 \begin{tabular}{lr}
 \vspace{-1.5cm}
\hspace{-1.5cm}
\\ \begin{subfigure}[b]{0.99\textwidth}
\includegraphics[width=0.49\textwidth]{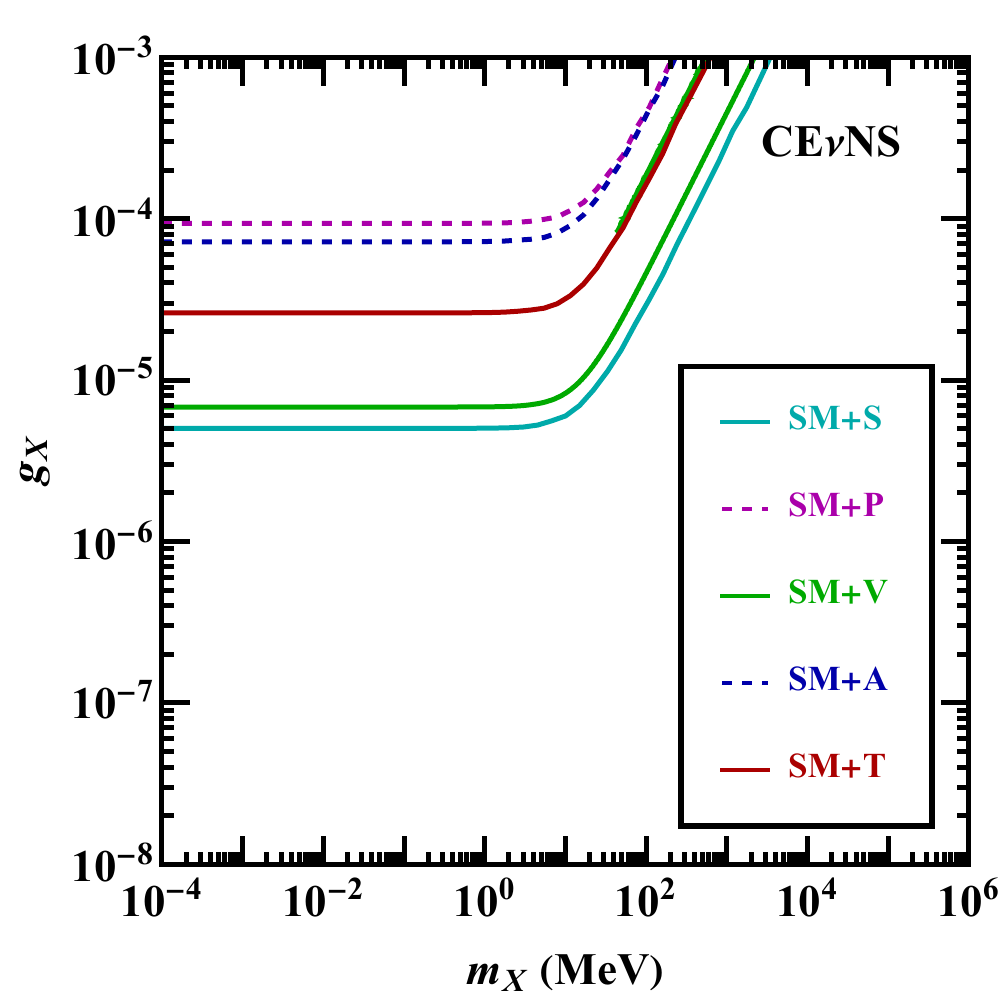}
\includegraphics[width=0.49\textwidth]{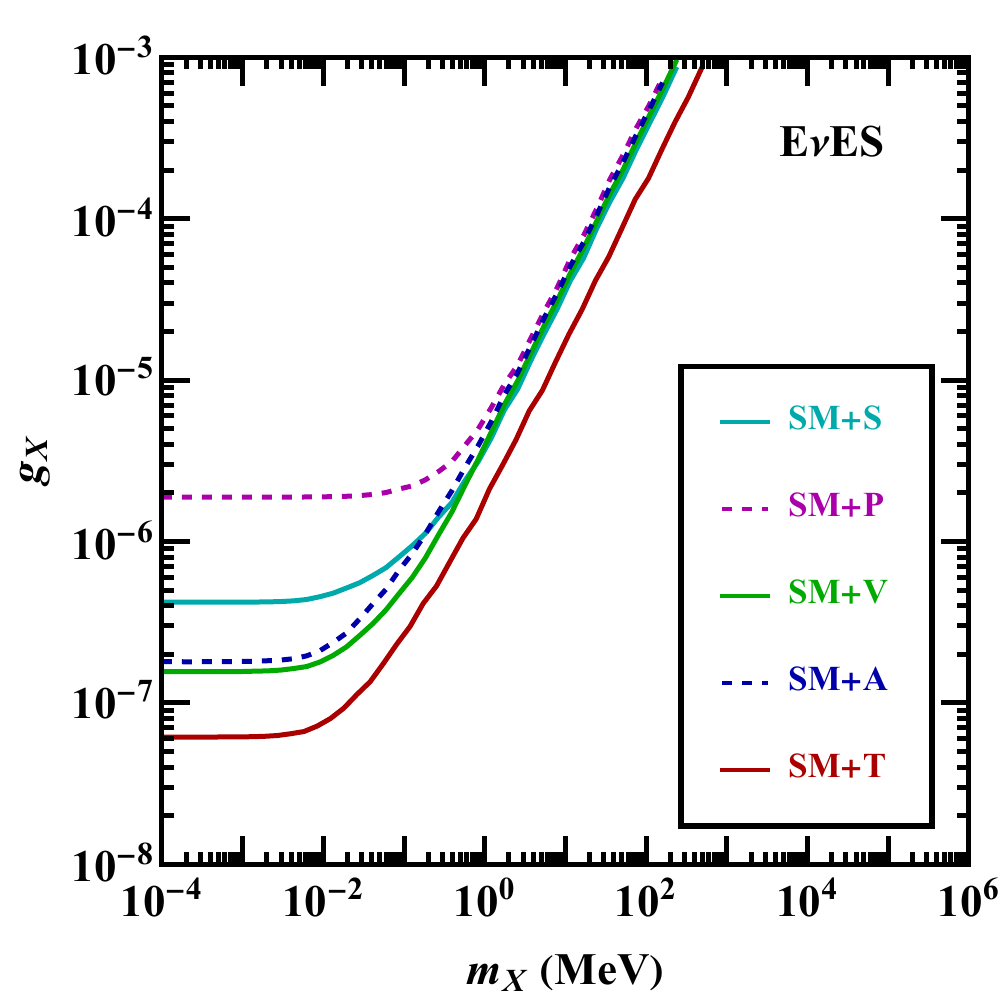}
\captionsetup{justification=centering}
\caption[a]{\textbf{Optimistic scenario}}
         \label{fig:bp2}
          \vspace{-0.01cm}
\end{subfigure}\\
\vspace{-0.5cm}

\begin{subfigure}[b]{0.99\textwidth}
\includegraphics[width=0.49\textwidth]{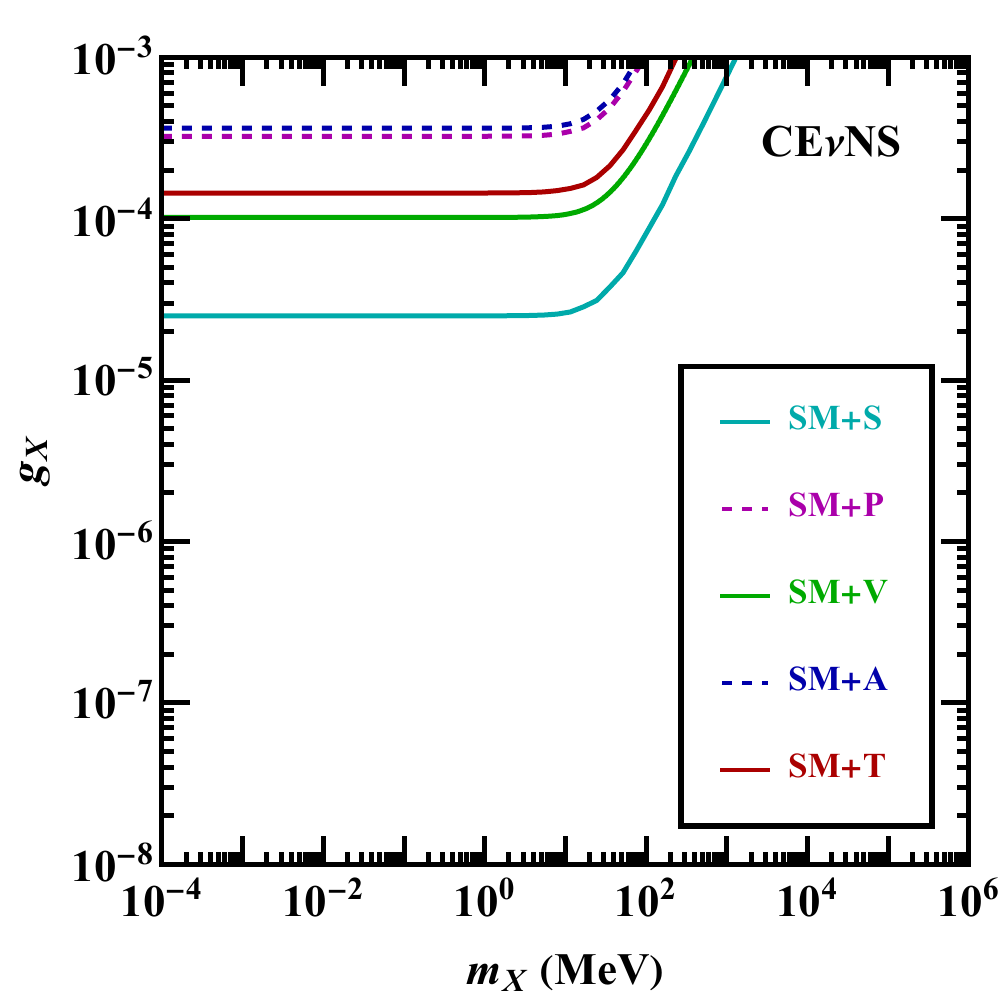}
\includegraphics[width=0.49\textwidth]{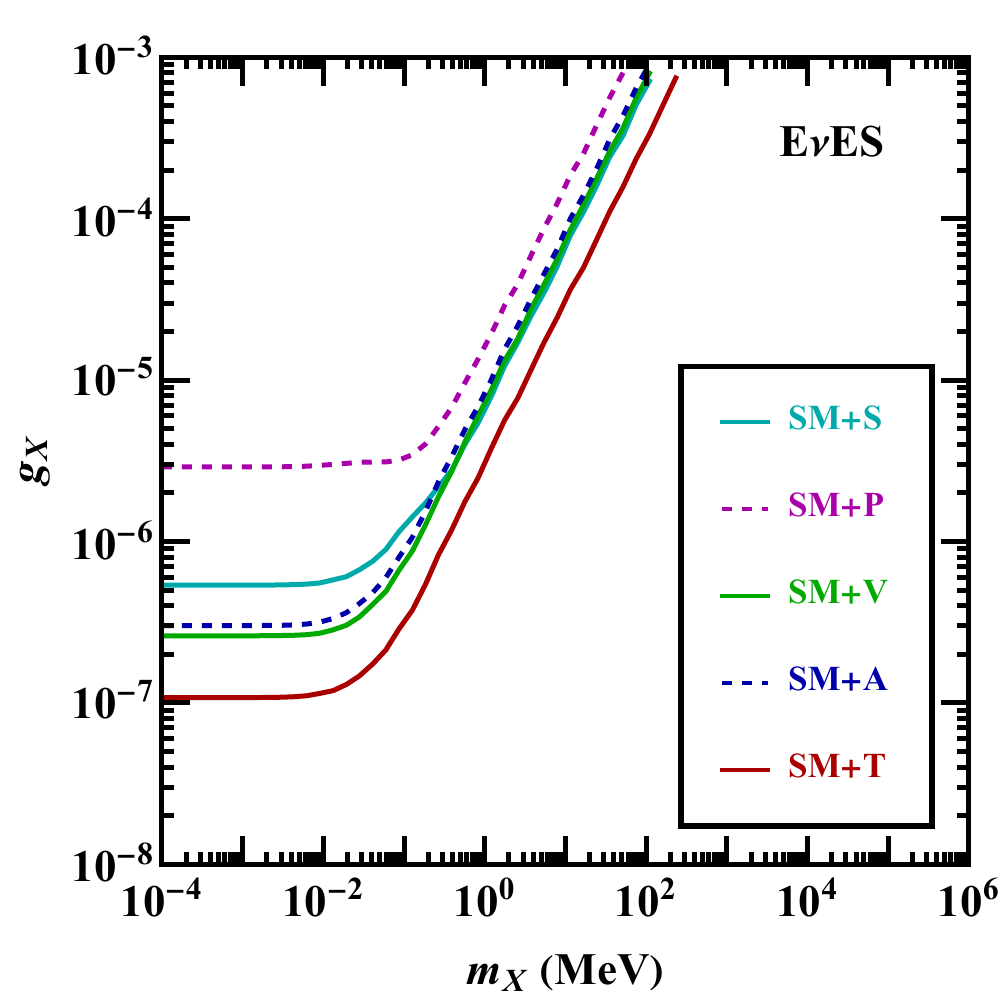}
\captionsetup{justification=centering}
\caption[a]{\textbf{Realistic scenario}}
         \label{fig:bp4}
\end{subfigure}\\ \\
\end{tabular}
 \end{center}
\vspace{-0.6cm}
\captionsetup{justification=raggedright}        
\caption{Projected sensitivities for the various $X= \{S, P, V, A, T\}$ interactions for \cevns (left) and \eves (right). The results are presented at 90\% C.L. Graphs correspond to the (a) optimistic and (b) realistic scenarios.}
\label{fig:sensitivities_X}
\end{figure}

\begin{figure}[!htbp]
\begin{center}
 \begin{tabular}{lr}
 \vspace{-1.5cm}
\hspace{-1.5cm}
\\ \begin{subfigure}[b]{0.99\textwidth}
\includegraphics[width=0.49\textwidth]{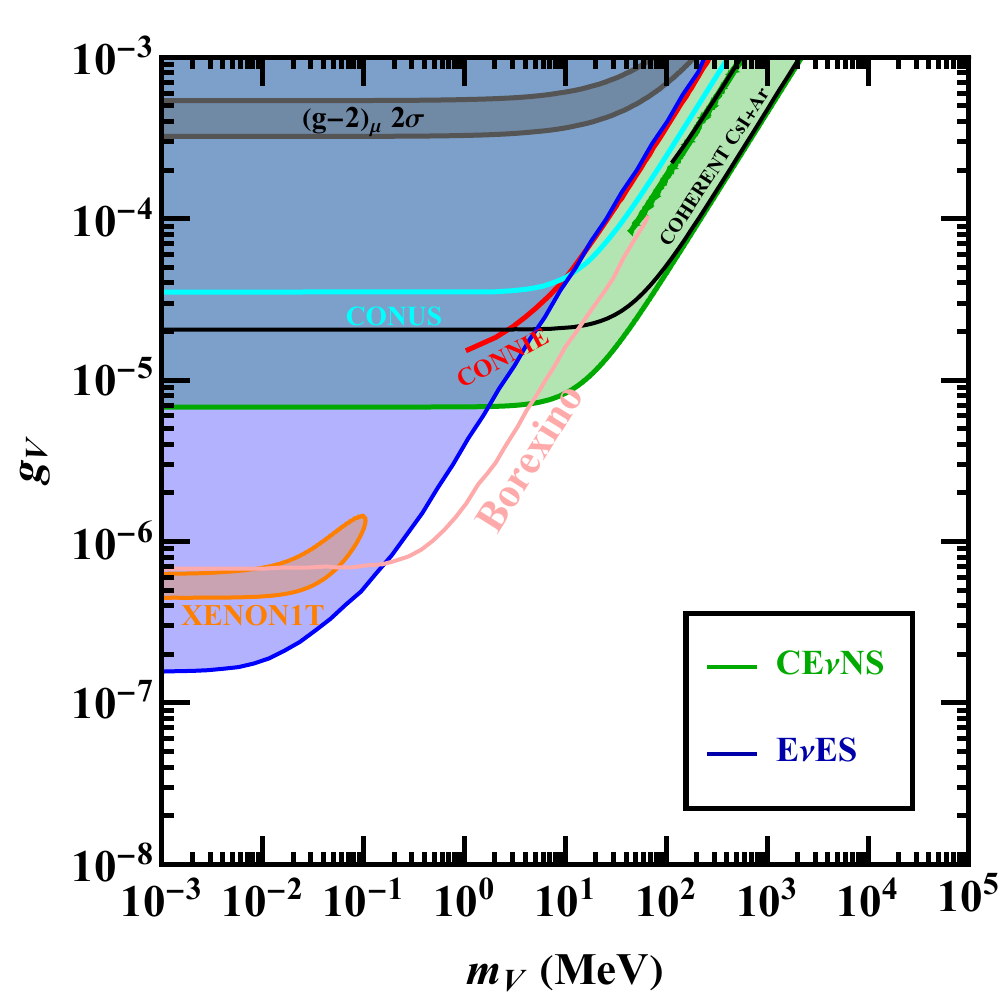}
\includegraphics[width=0.49\textwidth]{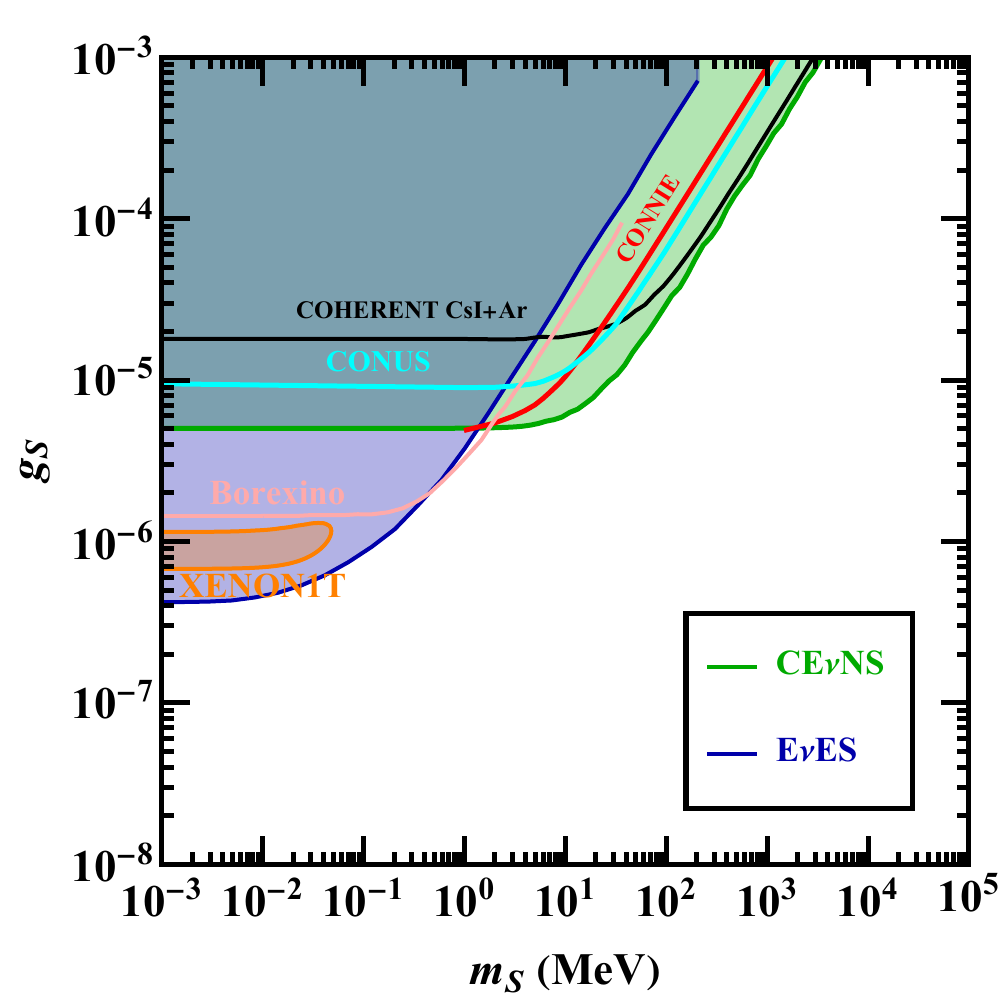}
\captionsetup{justification=centering}
\caption[a]{\textbf{Optimistic scenario:} For the universal vector (left) and the scalar (right) mediators.}
         \label{fig:bp2}
          \vspace{-0.01cm}
\end{subfigure}\\
\vspace{-0.5cm}

\begin{subfigure}[b]{0.99\textwidth}
\includegraphics[width=0.49\textwidth]{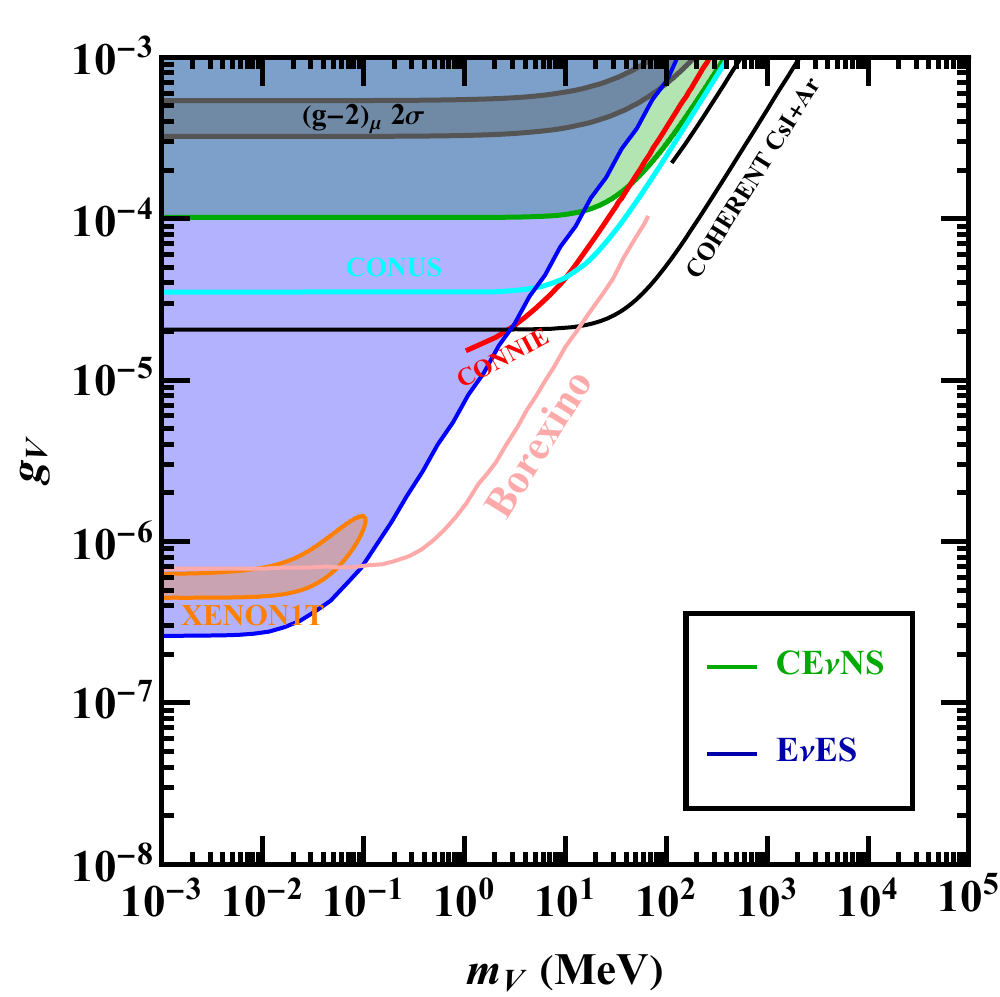}
\includegraphics[width=0.49\textwidth]{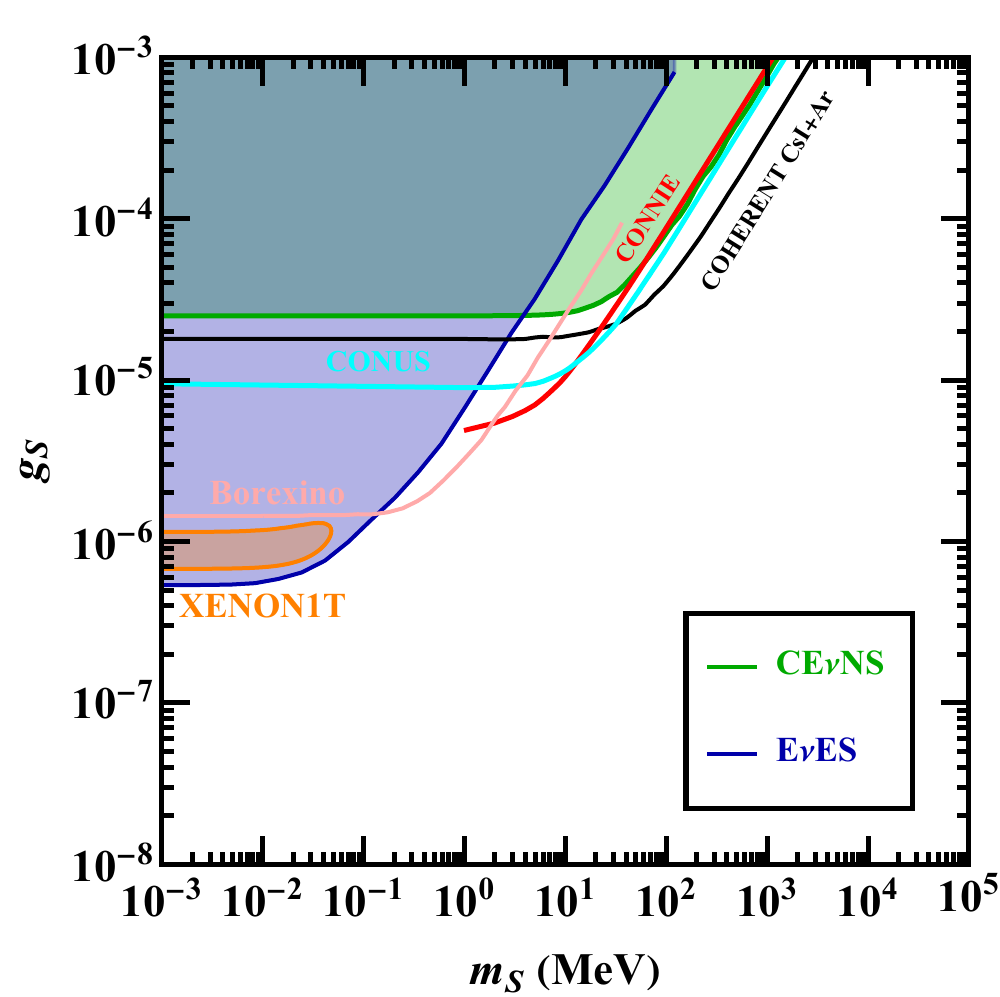}
\captionsetup{justification=centering}
\caption[a]{\textbf{Realistic scenario: }For the universal vector (left) and the scalar (right) mediators.}
         \label{fig:bp4}
\end{subfigure}\\ \\
\end{tabular}
 \end{center}
\vspace{-0.6cm}
\captionsetup{justification=raggedright}        
\caption{Projected sensitivity at 90\% C.L. in the parameter space $(m_V, g_V)$ for the universal vector mediator model (left) and in $(m_S, g_S)$ for a scalar mediator model (right). Graphs correspond to the (a) optimistic and (b) realistic scenarios. A comparison is given with existing constraints from dedicated \cevns experiments and XENON1T (see the text).}
\label{fig:zprime_universal}
\end{figure}

The resulting projected sensitivities in the ($m_X, g_X$) parameter space are illustrated at 90\% C.L. in the left and right panels of Fig.~\ref{fig:sensitivities_X}, respectively, while upper (lower) panels correspond to the optimistic (realistic) scenario. As can be seen, in a future CE$\nu$NS (E$\nu$ES) measurement among the different interaction channels the scalar (tensor) interaction will be constrained with maximum sensitivity, while  for both \cevns and \eves  the pseudoscalar interaction  will be the least constrained. A direct comparison of \cevns and \eves sensitivities leads to the conclusion that future \eves measurements will be a more powerful probe for the investigation of novel light mediators.

At this point it is interesting to compare our projected sensitivities with existing constraints in the literature. First, by focusing on the universal light vector mediator scenario, in the left panel of Fig.~\ref{fig:zprime_universal} we compare our present results at 90\% C.L.  with existing constraints from the combined analysis of COHERENT CsI+Ar data~\cite{Corona:2022wlb}, the recent CONNIE~\cite{CONNIE:2019xid} and CONUS~\cite{CONUS:2021dwh} data, the Borexino analysis of Ref.~\cite{Coloma:2022umy} as well as the anomalous magnetic moment of the muon reported in Ref.~\cite{Aoyama:2020ynm}. Also shown are the corresponding 90\% C.L. constraints coming from the XENON1T excess using E$\nu$ES calculated in this work. Our realistic scenario coincides with the XENON1T setup, hence the improvement on our E$\nu$ES-based constraints is driven by the factor $\sim 20$ larger exposure considered in the present analysis, while the shape difference is due to the XENON1T excess data (SM expectation) considered as $R_\text{exp}$ for the XENON1T (realistic scenario) analysis. The Borexino limits of Ref.~\cite{Coloma:2022umy} extend to larger mediator masses compared to our present results, since the analysis has been performed considering a larger recoil energy window, covering also CNO and $\mathrm{^8}$B neutrinos. By comparing the upper and lower graphs, we conclude that constraints from dedicated \cevns experiments will be overridden by the future \cevns measurements at direct detection dark matter detectors, given that the threshold and detection efficiencies will improve and the backgrounds will become even better understood. Finally, it becomes evident that the \eves channel dominates over \cevns in the low mass region for $m_V \leq 2 $MeV. 

Similar conclusions are drawn for the case of a scalar mediator, as shown in the right panel of Fig.~\ref{fig:zprime_universal}. Available results can be found from \cevns analyses in Refs.~\cite{CONNIE:2019xid,CONUS:2021dwh,Corona:2022wlb} and from the interpretation of XENON1T excess performed in this work (see also Refs.~\cite{Boehm:2020ltd,AristizabalSierra:2020edu}). Regarding the cases of $S, V, A$ interactions for both \cevns and \eves channels as well as for the case of $P$ using \eves only, results corresponding to the optimistic case can be found in Ref.~\cite{Cerdeno:2016sfi}. Let us note, however, that in the latter work the sensitivities were obtained by assuming different experimental configurations as well as by neglecting quenching and atomic effects for \cevns and \eves respectively. Regarding the cases of $P$ and $T$ interactions, preliminary results have been presented in Ref.~\cite{Demirci:2021zci} by  {analyzing} the COHERENT data. Again, the future sensitivities discussed here are about one order of magnitude more stringent than other studies. Finally, to the best of our knowledge, studies corresponding to our realistic case have not been done for any kind of  {model-dependent or -independent} light mediator.

\begin{figure}[t]
\begin{center}
 \begin{tabular}{lr}
 \vspace{-1.5cm}
\hspace{-1.5cm}
\\ \begin{subfigure}[b]{0.99\textwidth}
\includegraphics[width=0.49\textwidth]{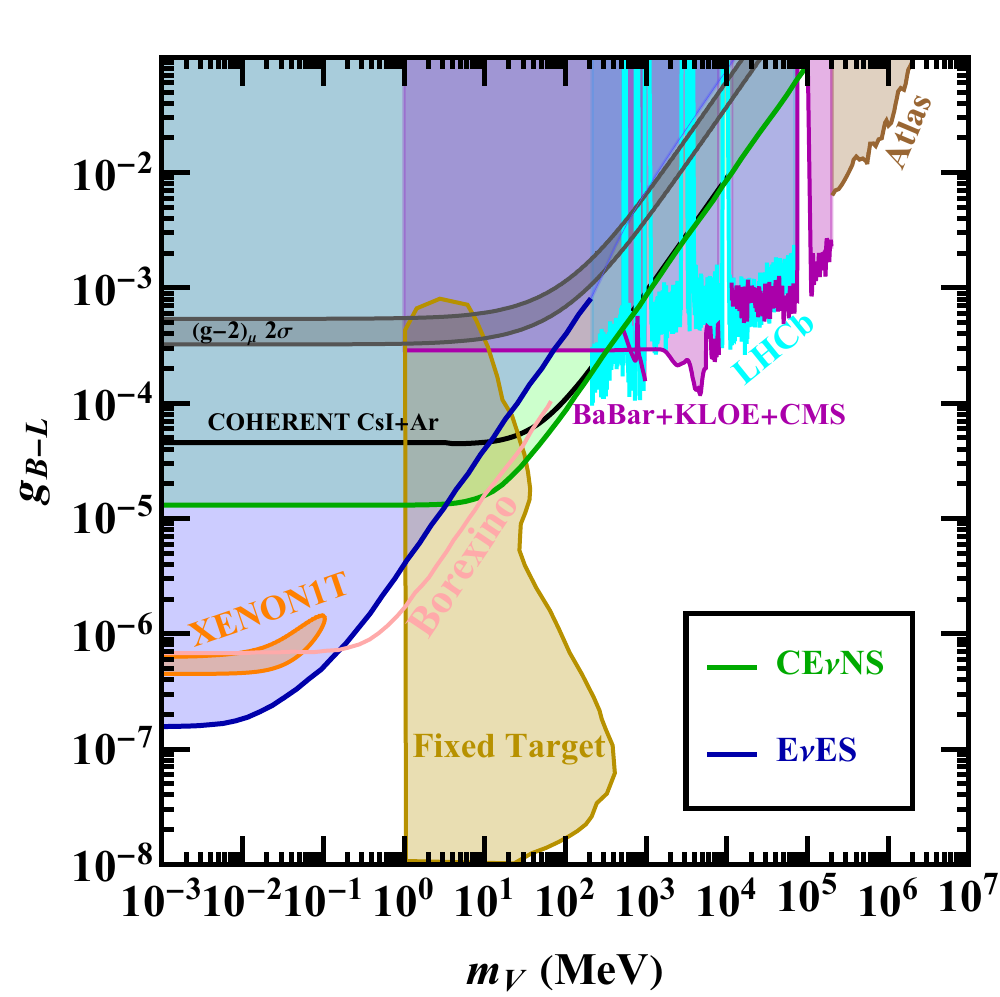}
\includegraphics[width=0.49\textwidth]{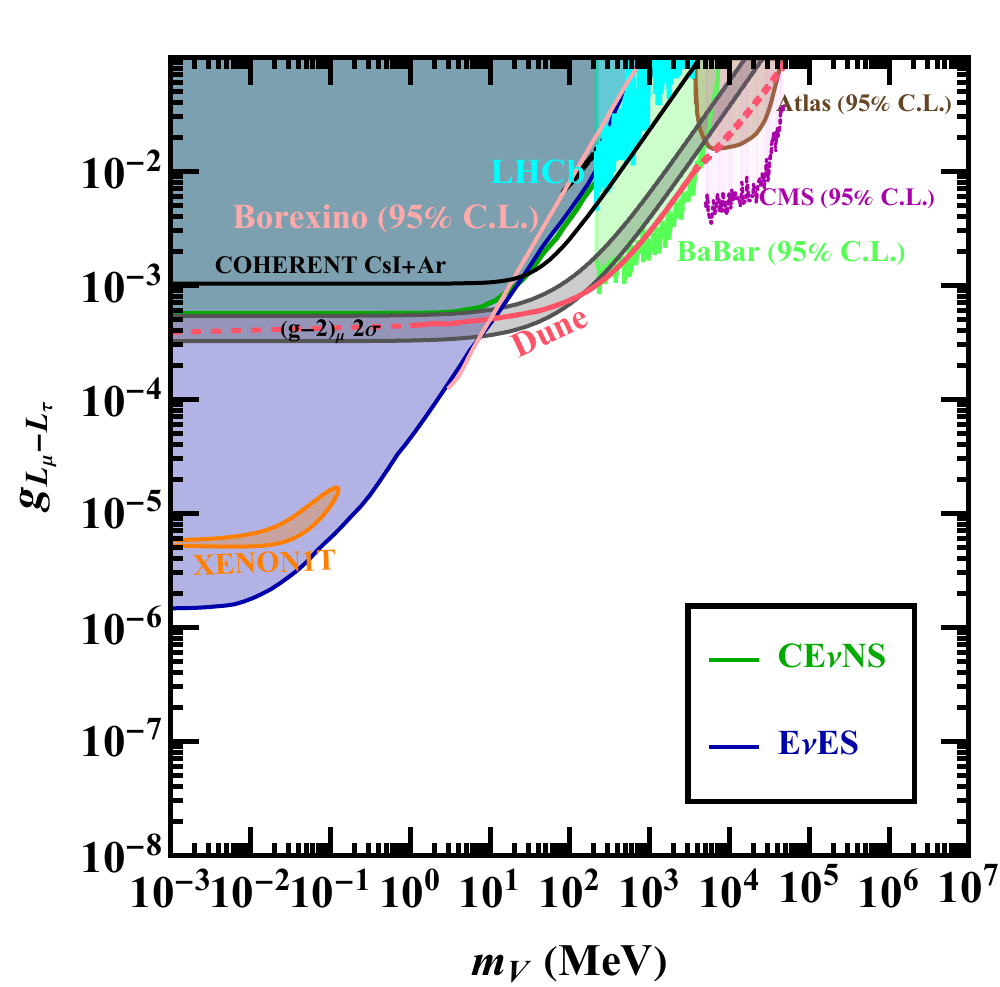}
\captionsetup{justification=centering}
\caption[a]{\textbf{Optimistic scenario:} For the $B-L$ (left)  and $L_\mu - L_\tau$ (right) models.}
         \label{fig:bp2}
          \vspace{-0.01cm}
\end{subfigure}\\
\vspace{-0.5cm}

\begin{subfigure}[b]{0.99\textwidth}
\includegraphics[width=0.49\textwidth]{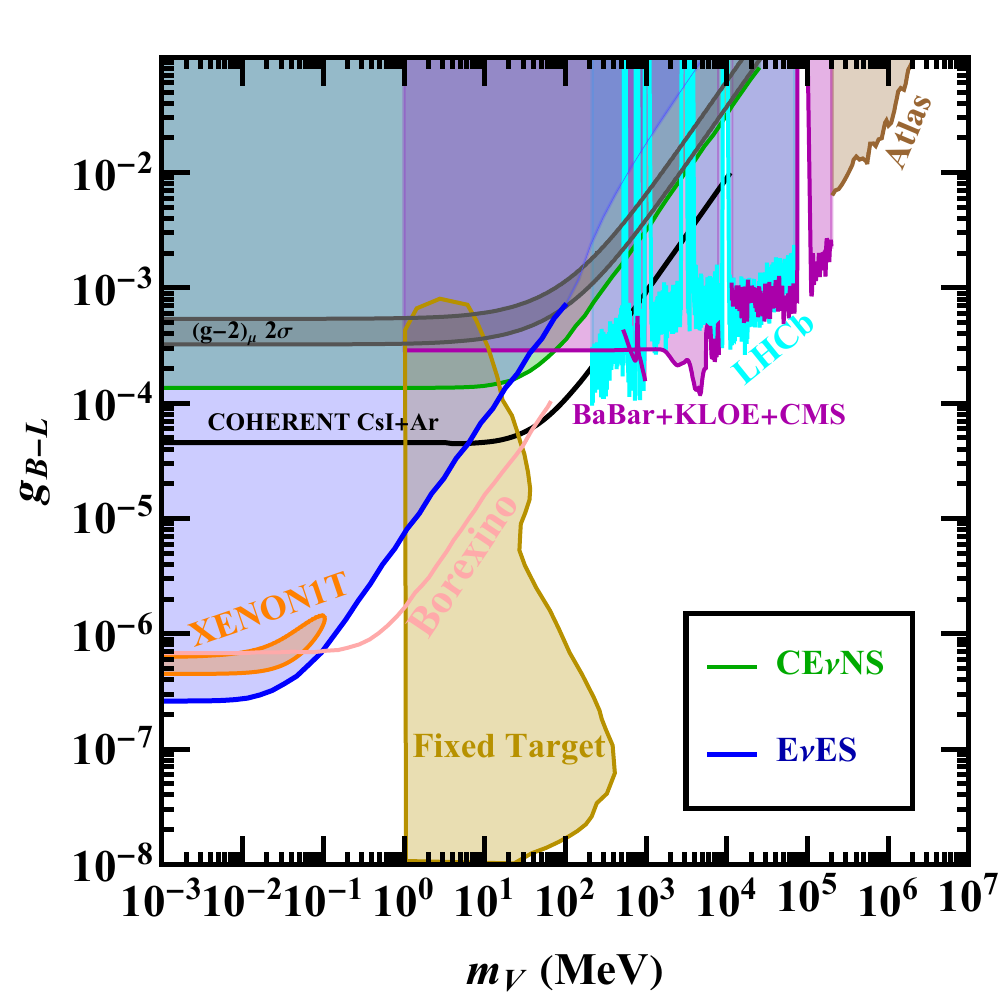}
\includegraphics[width=0.49\textwidth]{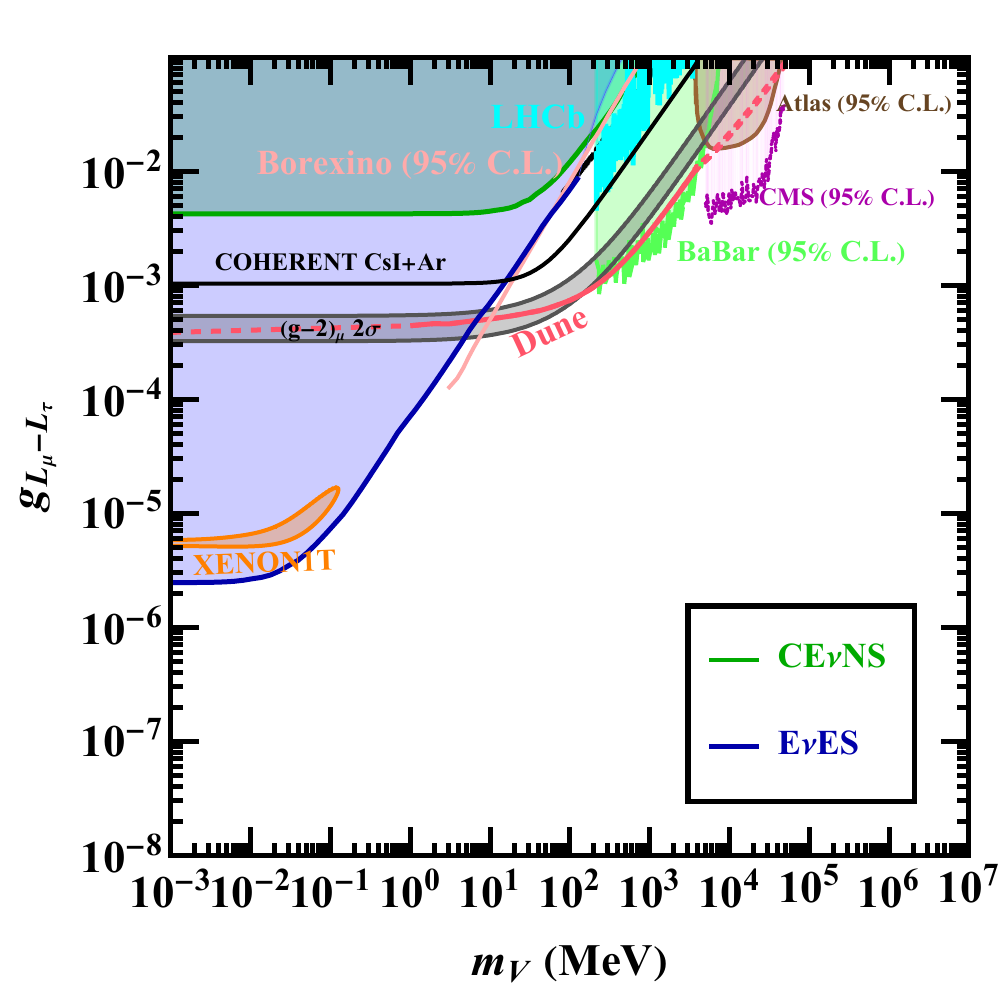}
\captionsetup{justification=centering}
\caption[a]{\textbf{Realistic scenario:} For the $B-L$ (left)  and $L_\mu - L_\tau$ (right) models.}
         \label{fig:bp4}
\end{subfigure}\\ \\
\end{tabular}
 \end{center}
\vspace{-0.6cm}
\captionsetup{justification=raggedright}        
\caption{Projected exclusion curves at 90\% C.L. obtained in the present work using \cevns and \eves for the $B-L$ (left) and $L_\mu - L_\tau$ (right) model. Graphs correspond to the (a) optimistic and (b) realistic scenarios. A comparison with relevant experimental constraints is also given (see the text).}
\label{fig:BL_Lmutau}
\end{figure}

We finally turn our attention to the vector mediator interactions predicted within the $U_{B-L}$ and $U_{L_\mu - L_\tau}$ gauge extensions. Before proceeding with the discussion of our statistical analysis, let us provide some clarifications regarding the calculation of the  number of \cevns and \eves events within the $L_\mu - L_\tau$ model. Since the $\nu_e - e^{-}$ coupling is vanishing,  we neglect the first term in Eq.(\ref{equn:EvES_Differential_Event_Rate}) and therefore a large portion of the solar neutrino flux will not contribute to the expected \eves rates. Likewise, in our CE$\nu$NS-based analysis  we consider only the oscillated $\nu_{\mu,\tau}$ fluxes of solar neutrinos, while from the atmospheric neutrino flux we consider only the relevant $\nu_\mu$ and  $\bar{\nu}_\mu$ components. For atmospheric neutrinos with sub-GeV neutrinos, matter-oscillation effects are negligible~\cite{Friedland:2004ah}. Following the procedure of Ref.~\cite{AristizabalSierra:2017joc}  one sees that the oscillated fluxes reaching the detector are: $\Phi^{\nu_e}_\text{atm} \approx \widetilde{\Phi}^{\nu_e}_\text{atm}$, $\Phi^{\nu_\mu}_\text{atm} \approx 2/3 \, \widetilde{\Phi}^{\nu_\mu}_\text{atm}$ and $\Phi^{\nu_\tau}_\text{atm} \approx 1/3 \, \widetilde{\Phi}^{\nu_\mu}_\text{atm}$, where $\widetilde{\Phi}^{\nu_\alpha}_\text{atm}$ denotes the unoscillated flux reported in Ref.~\cite{Battistoni:2005pd}. We have verified that our results remain unaffected when atmospheric neutrino oscillations are explicitly taken into account. Finally, regarding the DSN neutrino flux we assume that roughly $\Phi^{\nu_\mu + \bar{\nu}_\mu}_\text{DSN}  \approx \Phi^{\nu_\tau + \bar{\nu}_\tau}_\text{DSN}$. 

Left and right panels of Fig.~\ref{fig:BL_Lmutau} illustrate the projected sensitivity at 90\% C.L. for the studied $B-L$ and  $L_\mu - L_\tau$ model, respectively. The results are shown for both \cevns and \eves at a future dark matter direct detection experiment with the same general conclusions as discussed previously. Moreover, as expected, the exclusion curves corresponding to the $B-L$ model are more stringent. In order to compare with other experimental probes,  existing limits placed by 
dielectron resonances at ATLAS~\cite{Aaboud:2016cth}  {and} electron beam-dump fixed target experiments~\cite{Harnik:2012ni,Ilten:2018crw}, as well as Dark Photon searches at BaBar~\cite{Lees:2014xha,Lees:2017lec}, KLOE~\cite{KLOE-2:2018kqf}, CMS~\cite{CMS:2019kiy}, and LHCb~\cite{LHCb:2019vmc}\footnote{\scriptsize{BaBar, KLOE, CMS and LHCb limits, recasted to the $B-L$ case are obtained using the \href{https://gitlab.com/philten/darkcast}{Darkcast} software package.}}\normalsize, are superimposed. Further constraints from the \eves experiments TEXONO, GEMMA, BOREXINO, LSND and CHARM II  can be found in Ref.~\cite{Bilmis:2015lja}. Also shown are constraints derived from the analysis of the COHERENT data~\cite{Corona:2022wlb} and the XENON1T excess~\cite{AristizabalSierra:2020edu}. For the $L_\mu-L_\tau$ case, limits are available from recent analyses of Borexino data~\cite{Amaral:2020tga,Gninenko:2020xys}, from $4 \mu$ searches at BaBar~\cite{BaBar:2016sci}, CMS~\cite{CMS:2018yxg}, and ATLAS~\cite{Altmannshofer:2016jzy,ATLAS:2014jlg} as well as recasted limits from LHCb~\cite{LHCb:2019vmc}.
We furthermore show the corresponding sensitivity from the analysis of XENON1T excess performed in this work, by following the procedure of Ref.~\cite{Boehm:2020ltd}.
As can be seen, our projected sensitivities for \cevns dominate in  mass range $0.1 \leq m_V \leq 1$~GeV, being complementary to Babar and fixed target experiments. In the same vein, our projected sensitivities obtained using E$\nu$ES, are complementary to  fixed target experiments and particularly relevant for $m_V \leq 1$~MeV.

Before closing, we should stress that astrophysics and cosmology might place severe bounds to scalar and vector interactions~\cite{Farzan:2018gtr}. Those follow mostly from cosmological limits on the sum of neutrino masses, Supernova/stellar-cooling arguments  and sterile neutrino trapping as detailed in Refs.~\cite{AristizabalSierra:2020edu,AristizabalSierra:2019ykk} (see also Ref.~\cite{Das:2021nqj}). Given the large astrophysical uncertainties, such constraints should be considered as order of magnitude estimations while possible mechanisms to evade them are explained in Ref.~\cite{AristizabalSierra:2019ykk}.
Finally, it should be noted that upcoming experiments also have the potential to probe light mediator particles. However, our results provide about an order of magnitude more constrained bounds than those predicted for other future experiments such as neutrino trident interactions at DUNE~\cite{Altmannshofer:2019zhy},  {model-dependent} constraints for DARWIN~\cite{Amaral:2021rzw}, and constraints extracted from missing energy searches at NA64$\mu$~\cite{Amaral:2021rzw} (for expected limits from future beam dump experiments such as SHiP and FASER2 see Ref.~\cite{Das:2021nqj}). 

\section{Conclusions}
\label{sec:conclusions}

The new era of direct dark matter experiments with multiton mass scale and sub-keV operation threshold makes them favorable facilities with promising prospects for detecting astrophysical neutrino backgrounds to dark matter.
Prompted by the latter, we estimated the projected sensitivities to general Lorentz-invariant $X=\{S, P, V, A, T\}$ interactions through CE$\nu$NS- and E$\nu$ES-induced signals. With respect to light vector mediators, our study also considered the case of well-known $U_{B-L}$ and $U_{L_\mu - L_\tau}$ anomaly-free models.
To maximize the reliability of event rate simulations, important corrections from detector-specific quantities were taken into account such as the quenching factor and atomic binding effects for the case of \cevns and E$\nu$ES, respectively. 
Our statistical analysis was performed under the assumption of two benchmark scenarios, which allowed us to compare the maximum potential of a future direct dark matter detection experiment with the expected sensitivities when current detector specifications are explicitly accounted for. In the optimistic scenario we assumed a rather low  detection threshold of $0.1~\mathrm{keV_{ee}}$, which would correspond to an ionization-only analysis with a 100\% detection efficiency. In our goal to make our analysis as quantitative as possible, in the realistic scenario the effects of finite detection efficiency, nonflat backgrounds, and energy resolutions have been incorporated.
We furthermore considered  an exposure between $1\text{ and }20~\mathrm{ton \cdot yr}$ which is readily achievable by the current and future xenon detector technologies. 
Our present results imply that future \cevns or \eves measurements at a direct detection dark matter experiment will not only become sensitive to neutrinos coming (mainly) from the Sun, but will also offer competitive constraints to existing ones from dedicated \cevns and \eves experiments, if backgrounds are under control and sub-keV thresholds become possible. We have furthermore illustrated that the expected sensitivities will cover a large part of the parameter space, previously unexplored from collider probes and beam dump experiments, improving upon the existing bounds by about 1 order of magnitude. 

\acknowledgments

The authors are indebted to D. Aristizabal-Sierra for carefully reading the manuscript and for insightful comments. The authors also  acknowledge P. Mart\'inez-Mirav\'e, M. Williams, P. Ilten, K. Ni,  F. Gao and J. Pienaar for useful correspondence.
The research of D.K.P. is co-financed by Greece and the European Union (European Social Fund- ESF) through the Operational Programme ``Human Resources Development, Education and Lifelong Learning'' in the context of the project ``Reinforcement of Postdoctoral Researchers - 2nd Cycle'' (MIS-5033021), implemented by the State Scholarships Foundation (IKY). The work of RS is supported by the SERB, Government of India grant SRG/2020/002303.
%

%

\providecommand{\href}[2]{#2}\begingroup\raggedright\endgroup

\end{document}